\documentclass{article}
\usepackage[final,nonatbib]{neurips_2022}
\usepackage[utf8]{inputenc}
\usepackage{amsmath,amsthm,amssymb,caption,enumitem,subcaption,nicefrac}
\usepackage{thmtools,thm-restate}
\usepackage{bm}
\usepackage{algorithm}
\usepackage[noend]{algorithmic}
\usepackage[usenames,dvipsnames,table]{xcolor}
\usepackage{multirow}
\usepackage{caption}
\usepackage{subcaption}
\usepackage{hyperref}
\hypersetup{
	colorlinks=true,
	linkcolor=Blue,
	citecolor=black!30!OliveGreen,
	filecolor=Maroon,
	urlcolor=Maroon
}
\usepackage[capitalize,noabbrev,nameinlink]{cleveref}

\setlist[itemize]{label={\scriptsize$\blacktriangleright$}}

\newtheorem{theorem}{Theorem}

\newtheorem{lemma}[theorem]{Lemma}
\newtheorem{corollary}[theorem]{Corollary}
\newtheorem{fact}[theorem]{Fact}
\theoremstyle{definition}
\newtheorem{definition}[theorem]{Definition}

\newtheorem{observation}[theorem]{Observation}

\newcommand{\cX}{\mathcal{X}}
\newcommand{\cM}{\mathcal{M}}
\newcommand{\cR}{\mathcal{R}}
\newcommand{\cZ}{\mathcal{Z}}

\newcommand{\ind}{\mathbf{1}}
\newcommand{\eps}{\varepsilon}
\newcommand{\N}{\mathbb{N}}
\newcommand{\Z}{\mathbb{Z}}
\newcommand{\R}{\mathbb{R}}

\DeclareMathOperator{\E}{\mathbb{E}}
\newcommand{\cD}{\mathcal{D}}

\newcommand{\bh}{{\bm h}}
\newcommand{\bn}{{\bm n}}
\newcommand{\ban}{\bar{n}}
\newcommand{\tildeh}{\tilde{h}}
\newcommand{\tn}{\tilde{n}}
\newcommand{\tO}{\widetilde{O}}
\newcommand{\tOmega}{\widetilde{\Omega}}
\newcommand{\tbn}{{\bm \tn}}
\newcommand{\tbh}{{\bm \tildeh}}
\newcommand{\bbn}{\text{\boldmath $\ban$}}%

\renewcommand{\phi}{\varphi}
\newcommand{\bphi}{\bm{\phi}}
\newcommand{\hphi}{\hat{\phi}}
\newcommand{\hbphi}{\hat{\bphi}}

\newcommand{\red}{\mathrm{red}}
\newcommand{\hn}{\hat{n}}
\newcommand{\hf}{\hat{f}}

\newcommand{\bhn}{{\bm \hn}}
\DeclareMathOperator{\DLap}{DLap}

\DeclareMathOperator{\Var}{Var}

\DeclareMathOperator{\supp}{supp}

\newcommand{\set}[1]{\left \{ #1 \right \}}

\newcommand{\inabs}[1]{\left | #1 \right |}
\newcommand{\inparen}[1]{\left ( #1 \right )}
\newcommand{\insquare}[1]{\left [ #1 \right ]}

\definecolor{Gred}{RGB}{219, 50, 54}
\definecolor{Ggreen}{RGB}{60, 186, 84}
\definecolor{Gblue}{RGB}{72, 133, 237}
\definecolor{Gyellow}{RGB}{247, 178, 16}
\definecolor{ToCgreen}{RGB}{0, 128, 0}
\definecolor{myGold}{RGB}{231,141,20}
\definecolor{myBlue}{rgb}{0.19,0.41,.65}
\definecolor{myPurple}{RGB}{175,0,124}

\usepackage[colorinlistoftodos,prependcaption,textsize=scriptsize]{todonotes}
\providecommand{\Comments}{0}
\ifnum\Comments=1
\paperwidth=\dimexpr \paperwidth + 3.0cm\relax
\evensidemargin=\dimexpr\evensidemargin + 2.2cm\relax
\fi
\newcommand{\mytodo}[1]{\ifnum\Comments=1{#1}\fi}

\newcommand{\tableoftodos}{\ifnum\Comments=1 \listoftodos[Comments/To Do's] \fi}

\allowdisplaybreaks

\title{Anonymized Histograms\\
in  Intermediate Privacy Models}
\date{\today}

\begin{document}

\author{
Badih Ghazi
\hspace*{1cm}
\bf Pritish Kamath 
\hspace*{1cm}
\bf Ravi Kumar
\hspace*{1cm}
\bf Pasin Manurangsi
\\
Google Research\\
Mountain View, CA, US \\
\texttt{badihghazi@gmail.com,  pritish@alum.mit.edu,} \\
\texttt{ravi.k53@gmail.com, pasin@google.com}
}

\maketitle

\begin{abstract}
We study the problem of  privately computing the \emph{anonymized histogram} (a.k.a. \emph{unattributed histogram}), which is defined as the histogram without item labels. Previous works have provided algorithms with $\ell_1$- and $\ell_2^2$-errors of $O_\varepsilon(\sqrt{n})$ in the central model of differential privacy (DP).

In this work, we provide an algorithm with a nearly matching error guarantee of $\widetilde{O}_\varepsilon(\sqrt{n})$ in the shuffle DP and pan-private models. Our algorithm is very simple: it just post-processes the discrete Laplace-noised histogram!  Using this algorithm as a subroutine, we show applications in privately estimating symmetric properties of distributions such as entropy, support coverage, and support size.
\end{abstract}

\section{Introduction}

Computing histograms is among the most well-studied tasks in data analytics and machine learning. Suppose that there is a domain $[D] := \{1, \dots, D\}$, where the $i$th user's input is $z_i \in [D]$. The \emph{histogram} of the users' inputs $\{ z_1, \ldots, z_n \}$ is defined as $\bh := (h_1, \dots, h_D)$ where $h_j := |\{i \in [n] \mid z_i = j\}|$, i.e., the number of users who contribute item $j \in [D]$. For many tasks, however, the different items themselves are not important and it instead suffices to consider the \emph{anonymized histogram} (a.k.a. \emph{unattributed histogram}) corresponding to $\bh$, which is defined as $\bn_{\bh} := (n^{(1)}, \dots, n^{(D)})$ where $n^{(\ell)}$ denotes the $\ell$th largest element among the $h_j$'s. Whenever $\bh$ is clear from context, we will skip the subscript and denote the anonymized histogram as simply $\bn$.

Anonymized histograms have several applications including estimating symmetric properties of discrete distributions \cite{batu2000testing, AcharyaDOS17,CharikarSS19,HaoO19}, privately releasing the degree-distributions in social networks \cite{HayRMS10,HayLMJ09,raskhodnikova2015efficient}, and anonymizing password frequency lists \cite{BlockiDB16,bonneau2012science}. For more details, we refer the reader to \cite{Suresh19} and the references therein.

In this work, we study private anonymized histograms. The notion of privacy we study is differential privacy (DP) \cite{dwork2006calibrating, dwork2006our}, which has emerged as a very popular notion of private data analysis leading to numerous practical deployments \cite{erlingsson2014rappor,CNET2014Google, greenberg2016apple,dp2017learning, ding2017collecting, abowd2018us, LinkedINDP1, LinkedInDP2}.

Multiple works have studied the problem of computing private anonymized histogram, with the focus so far being on the central model of DP.\footnote{In the \emph{central model} of DP, a trusted curator has access to the raw inputs of the users, and is supposed to output a DP estimate of the desired function, in this case, the anonymized histogram.} Moreover, two measures of error have been studied: $\ell_1$-error and $\ell_2^2$-error\footnote{Defined as $\|\bn - \bhn\|_1 = \sum_{j \in [D]} |n^{(j)} - \hn^{(j)}|$ and $\|\bn - \bhn\|^2_2 = \sum_{j \in [D]} (n^{(j)} - \hn^{(j)})^2$, where $\bhn$ denotes the output anonymized histogram.}. For the $\ell_2^2$-error case, Hay et al.~\cite{HayRMS10} give an $\eps$-DP algorithm with an expected
error of $\tO(\sqrt{n} / \eps^2)$%
\footnote{In fact, Hay et al.~\cite{HayRMS10} proved a slightly stronger expected error bound of $O(d \log^3 n / \eps^2)$, where $d$ denotes the number of unique values appearing in $\bn$. Since there are at most $\sqrt{n}$ different values in the worst case, this guarantee yields an $O(\sqrt{n} \log^3 n / \eps^2)$ bound for any instance.}. As for the $\ell_1$-error, Blocki et al.~\cite{BlockiDB16} observed that the exponential mechanism yields an expected $\ell_1$-error of $O(\sqrt{n} / \eps)$ since there are at most $\exp(O(\sqrt{n}))$ anonymized histograms in total~\cite{HR18}; recently, this bound was improved to $O(\sqrt{n \log(1/\eps) / \eps})$ by Suresh~\cite{Suresh19}. On the lower bound front, Alda and Simon~\cite{AldaS18} proved a bound of $\Omega(\sqrt{n/\eps})$ for the expected $\ell_1$-error; recently, this was improved to $\Omega(\sqrt{n \log(1/\eps) / \eps})$ by Manurangsi~\cite{Manurangsi22}, matching the aforementioned upper bound of~\cite{Suresh19} to within a constant factor. The latter lower bound also applies to $(\eps, \delta)$-DP algorithms for any sufficiently small $\delta$ (depending only on $\eps$).

The %
anonymized histogram problem generalizes the {\sc Count-Distinct} problem, which asks for the number of items $j$ such that $h_j > 0$. {\sc Count-Distinct} can be easily solved in the central DP model by applying the discrete Laplace mechanism. In the (non-interactive) local DP setting, Chen et al.~\cite{ChenGKM21} proved a lower bound of $\Omega_{\eps}(n)$, which means that one cannot asymptotically beat the trivial algorithm that always outputs zero. The strong lower bounds on the error incurred by protocols in the local setting generally motivate the study of intermediate models of privacy including the pan-private \cite{DworkNPRY10} and shuffle  DP~\cite{bittau17, erlingsson2019amplification, CheuSUZZ19} models. In these models, it turns out that {\sc Count-Distinct} can be solved to within $\tO_{\eps}(\sqrt{n})$-error while lower bounds of $\Omega_\eps(\sqrt{n})$ are known~\cite{MirMNW11,BalcerCJM21}.

In this work, we show that, surprisingly, the anonymized histogram problem (which seems much harder than {\sc Count-Distinct}) can in fact be solved with essentially the same asymptotic error of $\tO_\eps(\sqrt{n})$ as {\sc Count-Distinct}, in both the pan-private and shuffle DP models.
On the other hand, the aforementioned lower bound~\cite{ChenGKM21} for {\sc Count-Distinct} also implies an $\Omega_\eps(n)$ lower bound for the expected $\ell_1$-error of anonymized histogram in the more challenging local DP model.
In other words, in the typical case where $\eps$ is an absolute constant, it is impossible to achieve any asymptotic advantage over the trivial algorithm that always outputs the all-zeros histogram.

\subsection{Our Results}

A prominent approach for computing private histograms in the central DP model is to add discrete\footnote{Our results can be adapted to continuous Laplace noise~\cite{dwork2006calibrating} with only a constant factor overhead in the error. We use discrete noise since it can be applied in the shuffle DP model (which is discrete in nature).} Laplace noise to each histogram entry. We show that there is a post-processing algorithm that takes such a noised histogram and produces an accurate estimate of the anonymized histogram:

\begin{theorem}[Informal; \Cref{thm:err-final,thm:err-l22}]\label{thm:informal-err-final}
There is an algorithm that takes in a noisy histogram, where an independent discrete Laplace noise of parameter $1/\eps$ is added to each entry, and outputs an approximate anonymized histogram such that the expected $\ell_1$- and $\ell^2_2$-errors are\footnote{$\tO(f)$  denotes $O(f \log^c f)$ for some constant $c > 0$ and $O_{\eps}(f)$ denotes $O(g(\eps) f)$ for some function $g(\cdot)$.} $\tO_\eps(\sqrt{n + D})$.
\end{theorem}

Note that there is a dependency of $\sqrt{D}$ in the error bound in \Cref{thm:informal-err-final}; when the domain size is large, this can dominate the $\sqrt{n}$ term. Fortunately, we show that this can be overcome by first randomly hashing into $B$ buckets before computing the histogram. By picking $B$ to be $O(n)$, we show that one can achieve an error that is $\tO_\eps(\sqrt{n})$ \emph{without} any dependency on $D$:

\begin{theorem}[Informal; \Cref{cor:hashing-err,thm:hashing-l22err}]
There is an algorithm that takes in a noisy hashed histogram, where an independent discrete Laplace noise of parameter $1/\eps$ is added to each bucket after hashing, and outputs an approximate anonymized histogram such that the expected $\ell_1$-error is $\tO_\eps(\sqrt{n})$. A similar algorithm holds, which uses two noisy hashed histograms and achieves expected $\ell^2_2$-error of $\tO_\eps(\sqrt{n})$.
\end{theorem}

Random hashing and computing discrete Laplace-noised histograms can be implemented in the pan-private\footnote{This is done by starting with $D$ i.i.d. discrete Laplace r.v.'s and incrementing each entry for the next item.} and shuffle DP settings~\cite{Ghazi0MP20,BalleBGN20}, where in the latter case we have to concede $\delta > 0$ in the privacy parameter. Thus, the theorem above yields:

\begin{corollary}
For any $\eps > 0, \delta \in (0, 1]$, there is an $\eps$-DP algorithm for anonymized histogram in the pan-private model and an $(\eps, \delta)$-DP algorithm in the shuffle DP model, with expected $\ell_1$- and $\ell_2^2$-errors of $\tO_\eps(\sqrt{n})$.%
\end{corollary}

As an immediate application of the above, we get algorithms for estimating symmetric properties of distributions; a distribution property is said to be \emph{symmetric} if it remains unchanged under relabeling of the domain symbols. For any (non-private) symmetric estimator with low sensitivity, we get a private estimator in the pan-private and shuffle DP models.

\begin{theorem}[Informal;\Cref{thm:symm-prop-estimation}]
For all $\eps > 0, \delta \in (0, 1]$, and distributions $\cD$, for any symmetric distribution property $f$, and any symmetric estimator $\hf$, there exists an $\eps$-DP mechanism $\cM$ in the pan-private model and an $(\eps,\delta)$-DP mechanism $\cM$ in the shuffle DP model, such that $\cM$ outputs an $\alpha$-approximation to $f(\cD)$ with high probability.  The sample complexity of the mechanism $\cM$  is given as $C_{\hf}(f, \alpha) + D_{\hf}(\alpha, \eps)$, where the first term is the non-private sample complexity of $\hf$ and the second term depends on the sensitivity of $\hf$.
\end{theorem}

In particular, in \Cref{sec:ml-applications}, we apply the above result to estimate the Shannon entropy, support coverage, and support size of discrete distributions, which, to the best of our knowledge, is the first such sample complexity bound in the pan-private and shuffle DP models.

\subsection{Overview of Techniques}
\label{sec:overview}

We now describe the high-level ideas of our algorithms and proofs.  For ease of exposition, we will occasionally be informal; all details are formalized in later sections. To describe our algorithm, we need definitions of prevalence and cumulative prevalence.
\begin{definition}[Prevalence and Cumulative Prevalence]
The \emph{prevalence} of a histogram $\bh$ is defined as $\bphi^{\bh} := (\phi^{\bh}_{0}, \dots, \phi^{\bh}_{n})$, where $\phi^{\bh}_{r} := |\{j \in [D] : h_{j} = r\}|$ is the number of entries with value $r$. The \emph{cumulative prevalence} of a histogram $\bh$ is defined as $\bphi^{\bh}_{\geq} := (\phi^{\bh}_{\geq 1}, \dots, \phi^{\bh}_{\geq n})$, where $\phi^{\bh}_{\geq r} := |\{j \in [D] : h_{j} \geq r\}|$ is the number of entries with value at least $r$.
\end{definition}
Prevalence and cumulative prevalence can be similarly defined for an anonymized histogram $\bn$; note that $\bphi^{\bh} = \bphi^{\bn_{\bh}}$ and $\bphi_{\ge}^{\bh} = \bphi_{\ge}^{\bn_{\bh}}$.
An important property of cumulative prevalence is that it preserves the $\ell_1$-distance.
\begin{observation}
For all anonymized histograms $\bn, \bhn$, it holds that
$\|\bphi^{\bn}_{\geq} - \bphi^{\bhn}_{\geq}\|_1 = \|\bn - \bhn\|_1$.
\end{observation}

We stress that, while cumulative prevalence has been used before in DP algorithms for computing anonymized histograms~\cite{Suresh19,Manurangsi22}, these algorithms require access to the true anonymized histogram first and therefore will only work in the central DP model.

\paragraph{\boldmath Algorithm for $\ell_1$-error.}
For each $j \in [D]$, using the discrete Laplace-noised count, we produce an unbiased estimate for whether $h_j \geq r$ for each $r \in [n]$. Adding these up over all $j \in [D]$ gives an unbiased estimate for $\phi^{\bh}_{\geq r}$. We then ``project'' the estimated $\bphi_{\geq}$ back so that it corresponds to a valid anonymized histogram. It can be seen that this last step can at most double the error.

While the algorithm described above is simple, it is unclear why it incurs an error of $\tO_\eps(\sqrt{n + D})$.  The analysis turns out to be quite delicate.  The key is that the unbiased estimator we use has variance that decreases exponentially with $|h_j - r|$. (In other words, the uncertainty is high only when $h_j$ is close to $r$.) Roughly speaking, this means that the total error is dominated by the error in the case where $h_j = r$. Suppose for simplicity that we only focus on this case.  Since there are $\phi^\bh_r$ entries satisfying the condition, the expected $\ell_1$-error for $\bphi_{\geq r}$ will be $\tO_\eps(\sqrt{\phi^\bh_r})$. Thus, in total, the $\ell_1$-error of the estimated cumulative prevalence is dominated by $\tO_\eps(\sum_{r \in [n]} \sqrt{\phi^\bh_r})$. We can now apply the Cauchy--Schwarz inequality to yield $\sum_{r \in [n]} \sqrt{\phi^\bh_{r}} \leq \sqrt{\sum_{r \in [n]}  1/r} \cdot \sqrt{\sum_{r \in [n]} r \cdot \phi^\bh_r} = \Theta(\sqrt{n \log n})$, where the last equality follows from the fact that  $\sum_{r \in [n]} r \cdot \phi^\bh_r$ is simply the total counts in the histogram.
This concludes our proof sketch.
The full proof can be found in \Cref{sec:post-process-dlap}.

\paragraph{Handling Large Domain Sizes.}
When $D \gg n$, we randomly hash into $B = \tO(n)$ buckets and compute the noisy ``reduced'' histogram on these $B$ buckets. We can use our approach above to compute the anonymized histogram on these $B$ buckets with $\ell_1$-error at most $\tO_{\eps}(\sqrt{n})$. While this is a reasonable approach, it does not yet give a good estimate for the original anonymized histogram: the reason is that there could be as many as $\tOmega(n)$ collisions due to hashing. To handle this, we define a function that ``inverts'' the reduced anonymized histogram to the original anonymized histogram. We then show that (i) this inverse has constant ``sensitivity'' and (ii) the reduced histogram is concentrated around its mean with an $\ell_1$-deviation of $\tO(\sqrt{n})$. Combining these two allows us to conclude that the inverse of the noisy anonymized histogram has expected $\ell_1$-error of $\tO_\eps(\sqrt{n})$ as desired. See \Cref{sec:large-domain} for details.

\paragraph{Privately Computing Symmetric Properties of Distributions.}
A large body of work starting with~\cite{AcharyaDOS17} has shown that plug-in estimators using the so-called profile maximum likelihood (PML) distribution achieve nearly optimal sample complexity for many symmetric distribution properties. At the core of these works is an important fact that many symmetric distribution properties have estimators (based on the anonymized histogram) with low sensitivity. Our algorithms then apply these estimators on our private anonymized histogram. The sensitivity of the estimator, together with the $\ell_1$-error bound we have shown for our anonymized histogram, immediately yield bounds on the errors of the estimators. See \Cref{sec:ml-applications} for details.

\paragraph{\boldmath Algorithm for $\ell^2_2$-error.}
Adapting the algorithm for $\ell^2_2$-error proceeds as follows: recall (e.g., by Hölder's inequality) that $\|\bn - \bhn\|_2^2 \leq \|\bn - \bhn\|_1 \cdot \|\bn - \bhn\|_\infty$. Notice also that the concentration of the noise implies that each discrete Laplace-noised count is within $O(\log D / \eps)$ of the true value. Due to this, we may only change the last (i.e., ``projection'') step by adding an extra constraint that each entry of the output estimated anonymized histogram is within $O(\log D / \eps)$ of the corresponding entry of the noisy histogram. This way we have ensured that $\|\bn - \bhn\|_\infty \leq O(\log D / \eps)$. Combining this with our earlier bound on the expected $\ell_1$-error immediately yields the desired bound on the $\ell_2^2$-error. See \Cref{apx:l22_error} for details.

\subsection{Other Related Work}

The original paper on the pan-private model~\cite{DworkNPRY10} also studies the problem of estimating $\phi^{\bn}_r$. However, their focus is on algorithms with small space complexity, and, if one were to sum up their error bounds for all $r$ directly, it would yield a trivial bound of $\Omega(n)$ on the total error in the anonymized histogram.

Several recent works have studied the testing/estimation/learning of symmetric and other properties of distributions under privacy constraints, mostly in the central DP model~\cite{AcharyaKSZ20, ASZ18, ADR17, CDK17, DHS15, KV18} and some in the local DP model~\cite{DJW17}.  In particular, 
Acharya et al.~\cite{AcharyaKSZ20} study privately computing symmetric distribution properties in the central model. Indeed, they also exploit the fact that the estimators have low sensitivity. However, in the central DP model, low sensitivity allows one to get a private estimate by adding Laplace noise directly to the non-private estimate, whereas we need to compute the estimator from our approximate anonymized histogram.

\section{Differential Privacy}

In this section, we review the basics of differential privacy (DP) and the shuffle DP and the pan-private models.  Let $[n]$ denote the set $\{1, \ldots, n\}$ and let $\ind[\cdot]$ denote the binary indicator function.

Two datasets $S = \set{z_1, \ldots, z_n}$ and $S' = \set{z_1', \ldots, z_n'}$ are said to be \emph{neighboring}, denoted $S \sim S'$, if there is an index $i \in [n]$ such that $z_j = z_j'$ for all $j \in [n] \setminus \set{i}$.  We recall the following definition~\cite{dwork2006calibrating,dwork2006our}:

\begin{definition}[Differential Privacy (DP)]
Let $\eps > 0$ and $\delta \in [0, 1]$.  
A randomized algorithm $\cM : \cZ^n \to \cR$ is \emph{$(\eps,\delta)$-differentially private} (\emph{$(\eps,\delta)$-DP}) if, for all $S \sim S'$ and all (measurable) outcomes $E \subseteq \cR$, we have that $\Pr[\cM(S)\in E] \le e^\eps \cdot \Pr[\cM(S')\in E] + \delta$. 
\end{definition}

We denote $(\eps, 0)$-DP as $\eps$-DP or \emph{pure}-DP. The case when $\delta > 0$ is referred to as \emph{approximate}-DP.  

In the \emph{central DP} model, all the inputs are stored and processed by an analyzer and the privacy is enforced only on the output of the analyzer.

\paragraph{Shuffle DP~\cite{bittau17, erlingsson2019amplification, CheuSUZZ19}.}
In the \emph{shuffle DP} model, there are three algorithms, namely, a local randomizer $\mathcal{R}$, a shuffler $\mathcal{S}$, and an analyzer $\mathcal{A}$.  Let $S = \{ z_1, \ldots, z_n \}$ be the input dataset.  The randomizer $\mathcal{R}$ takes $z \in S$ as input and outputs a multiset of messages.  The shuffler $\mathcal{S}$
takes the multiset of messages obtained from $\mathcal{R}$ applied to each $z \in S$ and permutes them randomly.  The analyzer $\mathcal{A}$ takes this permuted
multiset and computes the final output.  The privacy in the shuffle
model is enforced on the output of the shuffler $\mathcal{S}$, when a single input is changed.

\paragraph{Pan-privacy~\cite{DworkNPRY10}.} In the pan-private model, there is an algorithm that takes in a data stream of unbounded length consisting of elements in the domain. It is required that the internal state of the algorithm after any number of steps satisfies $\eps$-DP over the data stream prefix until that step.

We consider two data streams to be neighboring iff they differ by a single entry. This is sometimes referred to as the ``event-level'' or ``record-level'' setting~\cite{AminJM20}, in contrast to the ``user-level'' setting that was the focus of the original work on pan-privacy~\cite{DworkNPRY10}. We remark that the ``user-level'' setting is not appropriate for our notions of error as it allows for changing an entry of the histogram arbitrarily, meaning that we cannot achieve any non-trivial error guarantees. Furthermore, our definition above (which allows us to directly implement discrete Laplace-noised histogram) is different from the origin definition in~\cite{DworkNPRY10}, because the latter requires that revealing both the internal state and the final output \emph{simultaneously} are DP. Nonetheless, our algorithm can be easily adapted for this more restricted model, but with a worse dependency on $\eps$. We explain this in more detail in \Cref{sec:pan-privacy-extension}.

\paragraph{Histograms in Central DP.}
For a distribution $D$, let $x \sim D$ denote that the random variable $x$ is chosen from $D$.  For $p \in (0, 1)$, the \emph{discrete Laplace distribution} (aka symmetric geometric distribution), denoted by $\DLap(p)$, is the distribution supported on $\Z$ whose probability mass at $i \in \Z$ is $\frac{1-p}{1+p} \cdot p^{|i|}$. 

We use the following well-known fact about central DP and histograms.
\begin{fact}
\label{fac:hist-central}
The algorithm that adds $\DLap(e^{-\eps/2})$ noise to each entry of a histogram is $\eps$-DP in the central model.
\end{fact}
We refer to the output of this algorithm as the \emph{discrete Laplace-noised} histogram.

For a histogram $\bh$, we use $\bn_{\bh}$ to denote the anonymized histogram corresponding to $\bh$, often dropping the subscript whenever $\bh$ is clear from context.

 \section{Post-Processing Noised Histogram}
\label{sec:post-process-dlap}

In this section we describe our main algorithm, which obtains an anonymized histogram by suitably post-processing a noised histogram. We first define the following function $f: \Z \to \R$, which is used in our post-processing method described in \Cref{alg:L1-loss}.
\begin{align}
f(m) =
\begin{cases}
1 &\text{ if } m > 0, \\
1 + \frac{p}{(1 - p)^2} &\text{ if } m = 0, \\
-\frac{p}{(1 - p)^2} &\text{ if } m = -1, \\
0 &\text{ if } m < -1.
\end{cases}\label{eq:post-process-f}
\end{align}

\begin{algorithm}[h]
\caption{Anonymized Histogram Estimator w.r.t. $\ell_1$ loss.}
\label{alg:L1-loss}
\begin{algorithmic}
\STATE {\bf Input:} Discrete Laplace-noised histogram $\bh'$, i.e., $h'_j \sim h_j + \DLap(p)$\\[2mm]
\FOR{$r \in [n]$}
    \STATE $\hphi_{\geq r} \gets \sum_{j \in [D]} f(h'_j - r)$, where $f$ is defined in \eqref{eq:post-process-f}
\ENDFOR
\RETURN $\bhn$ that minimizes $\|\bphi^{\bhn}_{\geq} - \hbphi_{\geq}\|_1$.
\end{algorithmic}
\end{algorithm}

We give an efficient implementation of the above algorithm in \Cref{alg:L1-loss-fast} (\Cref{apx:L1-loss-fast}), which runs in time $O(D + n \log n)$. The main idea is to compute $\bhn$ that minimizes $\|\bphi^{\bhn}_{\geq} - \hbphi_{\geq}\|_1$ using $\ell_1$-isotonic regression.
We now state the main guarantee of the post-processing method.
\begin{theorem} \label{thm:err-final}
For all histograms $\bh$ with $\|\bh\|_1 = n$, the estimate $\bhn$ returned by \Cref{alg:L1-loss} satisfies
\[\textstyle \E[\|\bhn - \bn \|_1] ~\leq~ O\inparen{\sqrt{C_p (n + D)\log n}}\,.\] %
where, $C_p := \frac{p^2}{(1-p)^5} + \frac{p}{1-p}$.
\end{theorem}

From \Cref{fac:hist-central}, for $\eps$-DP, we may set $p = e^{-\eps/2}$. \Cref{thm:err-final} then gives a bound of $O(\sqrt{(n+D)\log n} / \eps^{2.5})$ on the expected $\ell_1$-error.

The crucial result needed in our analysis is the guarantee of the individual estimator $f$. We show that $f(h'_j - r)$ is an unbiased estimator of $\ind[h_j \geq r]$ and furthermore its variance decreases (exponentially) as $|h_j - r|$ increases.

\begin{lemma}%
\label{lem:ind-estimator}
For all $h, r \in \N \cup \{0\}$, if $h' \sim h + \DLap(p)$, then it holds that
\begin{align}
\E[f(h' - r)] &\textstyle~=~ \ind[h \geq r], \quad \text{and} \label{eq:unbiased-estimator}\\
\Var[f(h' - r)] &\textstyle~\leq~ 4 p^{|h - r|+1} \inparen{\frac{p}{(1 - p)^3} + (1 - p)}.\label{eq:estimator-variance}
\end{align}
\end{lemma}

\begin{proof}
Let $\tau := h - r, \tau' := h' - r$, and $x := p/(1-p)^2$.
Consider $g: \Z \to \R$ defined by $g(m) := f(m) - 1/2$. Notice that~\eqref{eq:unbiased-estimator} is equivalent to $\E[g(\tau')] = \ind[\tau \geq 0] - 1/2$. Due to symmetry, it suffices to consider the case where $\tau \geq 0$. In this case, we have
\begin{align*}
\E[g(\tau')] = \E_{Z \sim \DLap(p)}[g(\tau + Z)]
= \Pr[Z > -\tau] \cdot (1/2) + \Pr[Z \leq -\tau] \cdot \E[g(\tau + Z) \mid Z \leq -\tau].
\end{align*}
The last term can be expanded as follows:
\begin{align*}
&\E[g(\tau + Z) \mid Z \leq -\tau] \\
&= (1/2 + x) \cdot \Pr[Z = -\tau \mid Z \leq -\tau] - (1/2 + x) \cdot \Pr[Z = -\tau - 1 \mid Z \leq -\tau] \\
&\quad - (1/2) \cdot \Pr[Z < -\tau - 1 \mid Z \leq -\tau] \\
&= (1/2 + x) \cdot (1 - p) - (1/2 + x) \cdot p(1 - p) - (1/2) \cdot p^2 %
= 1/2.
\end{align*}

Combining the two equations above, we arrive at $\E[g(\tau')] = \Pr[Z > -\tau] \cdot (1/2) + \Pr[Z \leq -\tau] \cdot (1/2) = 1/2$, thereby proving \eqref{eq:unbiased-estimator}.

To prove~\eqref{eq:estimator-variance}, notice again that $\Var[f(\tau')] = \Var[g(\tau')]$. Again, due to symmetry, we may only consider the case $\tau \geq 0$. Here, we have
\begin{align*}
\Var[g(\tau')] &~=~ \E_{Z \sim \DLap(p)}[(g(\tau + Z) - 1/2)^2] \\
&~=~  \Pr[Z \leq -\tau] \cdot \E[(g(\tau + Z) - 1/2)^2 \mid Z \leq -\tau] \\
&~\leq~ p^{\tau} \cdot \E[(g(\tau + Z) - 1/2)^2 \mid Z \leq -\tau].
\end{align*}
Similar to before, we can expand the last term as
\begin{align*}
&\E[(g(\tau + Z) - 1/2)^2 \mid Z \leq -\tau] \\
&= x^2 \cdot \Pr[Z = -\tau \mid Z \leq -\tau] + (1 + x)^2 \cdot \Pr[Z = -\tau - 1 \mid Z \leq -\tau] \\
&\quad + 1 \cdot \Pr[Z < -\tau - 1 \mid Z \leq -\tau] \\
&= x^2 \cdot (1-p) + (1+x)^2 \cdot p(1 - p) + p^2 \\
&\leq p^2 / (1 - p)^3 + 2(1 + x^2) \cdot p(1 - p) + p^2 \\
&\leq 4 p^2 / (1 - p)^3 +  2p(1 - p).
\end{align*}
Plugging this into the above, we get $\Var[g(\tau')] \leq 4 p^{\tau+1} (p/(1 - p)^3 + (1 - p))$ as desired.
\end{proof}

The following is an immediate consequence of~\Cref{lem:ind-estimator}, by summing over all $j \in [D]$.

\begin{observation} \label{obs:cumulative-variance}
For all $r \in [n]$, and $\kappa := 4p\inparen{\frac{p}{(1 - p)^3} + (1 - p)}$ it holds that
\[ \textstyle \E[\hphi_{\geq r}] ~=~ \phi^{\bn}_{\geq r} \quad \text{and} \quad \Var[\hphi_{\geq r}] ~\leq~ \sum_{\ell=0}^n \kappa \cdot p^{|\ell - r|} \cdot \phi^{\bn}_{\ell}\,. \]
\end{observation}

\begin{proof}[Proof of \Cref{thm:err-final}]
The expected $\ell_1$-error is bounded as
\begin{align}
\|\bhn - \bn \|_1 &~=~ \|\bphi^{\bhn}_{\ge} - \bphi^{\bn}_{\ge} \|_1 ~\le~ \|\bphi^{\bhn}_{\ge} - \hbphi_{\geq}\|_1 + \|\hbphi_{\geq} - \bphi^{\bn}_{\ge} \|_1 ~\le~ 2 \cdot \|\bphi^{\bhn}_{\ge} - \hbphi_{\geq}\|_1 \nonumber\\
\Longrightarrow~\|\bhn - \bn \|_1 &\textstyle~\le~ 2 \cdot \sum_{r \in [n]} |\phi^{\hn}_{\geq r} - \hphi_{\geq r}|, \label{eq:err-final-1}
\end{align}
where we use that $\|\bphi^{\bhn}_{\ge} - \hbphi_{\geq}\|_1 \le \|\hbphi_{\geq} - \bphi^{\bn}_{\ge} \|_1$ by our choice of $\bhn$.
From \Cref{obs:cumulative-variance}, we have
\begin{align*}
\E\left[|\phi^{\hn}_{\geq r} - \hphi_{\geq r}|\right] 
~\leq~ \sqrt{\Var[\hphi_{\geq r}]}
&\textstyle~\le~\sqrt{\kappa} \cdot \sqrt{\sum_{\ell=0}^n p^{|\ell - r|} \cdot \phi^{\bn}_{\ell}}.
\end{align*}
Combining this with \eqref{eq:err-final-1}, we have
\begin{align*}
\E[\|\bhn - \bn\|_1]
&\textstyle~\leq~ 2 \cdot \sum_{r \in [n]} \E\insquare{|\phi^{\hn}_{\geq r} - \hphi_{\geq r}|}\\
&\textstyle~\leq~ 2\sqrt{\kappa} \cdot \left(\sum_{r \in [n]} \sqrt{\sum_{\ell=0}^n p^{|\ell - r|} \cdot \phi^{\bn}_{\ell}}\right) \\
&\textstyle~\leq~ 2\sqrt{\kappa} \cdot \sqrt{\sum_{r \in [n]} \frac{1}{r}} \cdot \sqrt{\sum_{r \in [n]} r \cdot \left(\sum_{\ell=0}^n p^{|\ell - r|} \cdot \phi^{\bn}_{\ell}\right)} \quad \text{(Cauchy--Schwarz)}\\
&\textstyle~\leq~ \sqrt{\kappa} \cdot O(\sqrt{\log n}) \cdot \sqrt{\sum_{\ell=0}^n \phi^{\bn}_\ell \cdot \left(\sum_{r \in [n]} r \cdot p^{|\ell - r|}\right)} \\
&\textstyle~\leq~ \sqrt{\kappa} \cdot O(\sqrt{\log n}) \cdot \sqrt{\sum_{\ell=0}^n \phi^{\bn}_\ell \cdot 2(\ell+1) \cdot \left(\sum_{t=0}^{\infty} (t + 1) \cdot p^{t}\right)} \\
&\textstyle~=~ \sqrt{\kappa} \cdot O(\sqrt{\log n}) \cdot \sqrt{\sum_{\ell=0}^n \phi^{\bn}_\ell \cdot 2(\ell+1) \cdot \left(1/(1 - p)^2\right)} \\
&\textstyle~=~ \sqrt{\kappa/(1 - p)^2} \cdot O(\sqrt{\log n}) \cdot \sqrt{\left(\sum_{\ell=0}^n \ell \cdot \phi^{\bn}_\ell\right) + \left(\sum_{\ell=0}^n \phi^{\bn}_\ell\right)} \\
&\textstyle~=~ \sqrt{\kappa/(1 - p)^2} \cdot O(\sqrt{\log n}) \cdot \sqrt{n + D}. \qedhere
\end{align*}
\end{proof}

\section{Reducing Domain Size via Hashing} \label{sec:large-domain}

In this section we propose an algorithm to handle the case where $D \gg n$.  The approach in this case is to hash the domain into something smaller.  Let $B \in \N$ be the number of hash values. The distribution of the anonymized histogram produced after random hashing into $B$ buckets is equivalent to the %
following process:
\begin{itemize}[leftmargin=5mm,nosep]
\item Let $\bn = (n^{(1)}, \dots, n^{(D)})$ be the starting anonymized histogram.
\item Pick a uniformly random hash function $H: [D] \to [B]$.
\item Let $\bh^{\red} := (h^{\red}_1, \dots, h^{\red}_B)$ denote the reduced histogram given by $h^{\red}_i = \sum_{j \in H^{-1}(i)} n^{(j)}$.
\item Let $\bn^{\red}$ denote the corresponding anonymized histogram.
\end{itemize}

Let $\Gamma^B$ be the mapping from $\bn$ to $\E[\bphi_{\geq}^{\bn^{\red}}]$ where $\bn^{\red}$ is generated as above, and the expectation is over the choice of random hash functions $H$. %
With this notation, we present  \Cref{alg:hash-L1-loss}. %

\begin{algorithm}[h]
\caption{Anonymized Histogram Estimator w.r.t. $\ell_1$ loss, for large domains.}
\label{alg:hash-L1-loss}
\begin{algorithmic}
\STATE {\bf Input:} Discrete Laplace-noised histogram $\tilde{\bh}^{\red}$, i.e., $\tildeh^{\red}_j \sim h^{\red}_j + \DLap(p)$\\[2mm]
\STATE Compute an estimate $\bhn^{\red}$ of $\bn^{\red}$ using \Cref{alg:L1-loss}
\RETURN $\bhn$ that minimizes $\|\Gamma^B(\bhn) - \bphi_{\geq}^{\bhn^{\red}}\|_1$, s.t. $\|\bhn\|_1 = n$
\end{algorithmic}
\end{algorithm}

We give an efficient implementation of (a variant of) the above algorithm in \Cref{alg:hash-L1-loss-approx-fast-simplified} (\Cref{apx:hash-L1-loss-fast}), which runs in time $\tO_{\eps}(D + n \log n)$. The main result of this section is the following:
\begin{theorem} \label{thm:hashing-err}
For all $B > 4n$ satisfies
\begin{align*}
    \E[\|\bhn - \bn\|_1] &~\leq~ O(\E[\|\bhn^{\red} - \bn^{\red}\|_1] + n \log n / \sqrt{B})\\
    &~\leq~ O\inparen{\sqrt{C_p (n+B) \log n} + n \log n / \sqrt{B}}\qquad \text{(using \Cref{thm:err-final}).}
\end{align*}
\end{theorem}

By setting $B = n \sqrt{\log n}$, we get the following corollary.
\begin{corollary}\label{cor:hashing-err}
For all $0 < \eps \leq O(1)$, \Cref{alg:hash-L1-loss} for $p = e^{-\eps/2}$ and $B = n \sqrt{\log n}$ is an $\eps$-DP algorithm, and achieves an expected $\ell_1$-error of $E(n, \eps) = O(\sqrt{n} (\log n)^{3/4} / \eps^{2.5})$.%
\footnote{By choosing $B = n \sqrt{\eps^{2.5} \log n}$, the bound in \Cref{cor:hashing-err} could be improved to $O(\sqrt{n} ((\log n)^{3/4} / \eps^{1.25} + (\log n)^{1/2} / \eps^{2.5}))$ for $\eps \geq \Omega(\log^{-0.4} n)$; we state the simpler bound for brevity.}
\end{corollary}

We describe the main steps in the proof of \Cref{thm:hashing-err}.

\paragraph{\boldmath Lipschitzness of Inverse of $\Gamma^B$.}
We start by showing that the ``inverse'' of $\Gamma^B$ is $O(1)$-Lipschitz:

\begin{restatable}{lemma}{inverselipschitz}[Proof in \Cref{apx:inverse-lipschitz-proof}] \label{lem:inverse-lipschitz}
For $B > 4n$ and all anonymized histograms $\bn, \bn'$ with $\|\bn\|_1, \|\bn'\|_1 \le n$,
\[\|\Gamma^B(\bn) - \Gamma^B(\bn')\|_1 \geq \|\bn - \bn'\|_1 / 4\,.\]
\end{restatable}

\paragraph{\boldmath Concentration of $\bn^{\red}$.}
We now bound the expected $\ell_1$-distance between $\bphi_{\geq}^{\bn^{\red}}$ and its expectation $\Gamma^B(\bn)$.

\begin{lemma} \label{lem:hash-concen}
Assume that $B \geq 2n$. Then, $\E[\|\bphi_{\geq}^{\bn^{\red}} - \Gamma^B(\bn)\|_1] \leq O(n \log n / \sqrt{B})$.
\end{lemma}

\begin{proof}
We have
\begin{align}
\E[\|\bphi_{\geq}^{\bn^{\red}} - \Gamma^B(\bn)\|_1]
&\textstyle~=~ \sum_{r\in[n]} \E[|\phi_{\geq r}^{\bn^{\red}} - \Gamma^B(\bn)_r|] \nonumber\\
&\textstyle~=~ \sum_{r\in[n]} \E[|\phi_{\geq r}^{\bn^{\red}} - \E[\phi_{\geq r}^{\bn^{\red}}]|]\nonumber\\
&\textstyle~\leq~ \sum_{r \in [n]} \sqrt{\Var[\phi_{\geq r}^{\bn^{\red}}]}.\label{eq:hash-concern-1}
\end{align}

We next use the following bound on the variance. 
\begin{restatable}{lemma}{varphihashedterm}[Proof in \Cref{apx:var-phi-hashed-term-proof}] \label{lem:var-phi-hashed-term}
For all $r \in [n]$, $\Var[\phi_{\geq r}^{\bn^{\red}}] \leq \frac{16n}{B} \cdot \left(\frac{n}{r^2} + \sum_{t \in [r - 1]} \frac{t \cdot \phi^{\bn}_t}{r(r - t)}\right)$.
\end{restatable}

Plugging \Cref{lem:var-phi-hashed-term} into \eqref{eq:hash-concern-1}, we have
\begin{align*}
\E[\|\bphi_{\geq}^{\bn^{\red}} - \Gamma^B(\bn)\|_1] 
&\textstyle~\leq~ O\left(\sqrt{n/B}\right) \cdot \sum_{r \in [n]} \sqrt{\frac{n}{r^2} + \sum_{t \in [r - 1]} \frac{t \cdot \phi^{\bn}_t}{r(r - t)}} \\
\text{(Cauchy--Schwarz)} \quad &\textstyle~\le~ O\left(\sqrt{n/B}\right) \cdot \sqrt{\sum_{r \in [n]} \frac{1}{r}} \sqrt{\sum_{r \in [n]} r \cdot \left(\frac{n}{r^2} + \sum_{t \in [r - 1]} \frac{t \cdot \phi^{\bn}_t}{r(r - t)}\right)} \\
&\textstyle~=~ O\left(\sqrt{n \log(n) /B}\right) \sqrt{\sum_{r \in [n]}  \left(\frac{n}{r} + \sum_{t \in [r - 1]} \frac{t \cdot \phi^{\bn}_t}{r - t}\right)} \\
&\textstyle~=~ O\left(\sqrt{n \log(n) /B}\right) \sqrt{O(n \log n) + \sum_{t \in [n - 1]} \sum_{\ell \in [n - t]} \frac{t \cdot \phi^{\bn}_t}{\ell}} \\
&\textstyle~\leq~ O\left(\sqrt{n \log(n) /B}\right) \sqrt{O(n \log n) + O(\log n) \cdot \sum_{t \in [n - 1]} {t \cdot \phi^{\bn}_t}} \\
&\textstyle~=~ O\left(\sqrt{n \log(n) /B}\right) \sqrt{O(n \log n)} \\
&\textstyle~=~ O(n \log n / \sqrt{B})\,.\qedhere
\end{align*}
\end{proof}

\paragraph{Putting things together.}
With all the components ready, we can now prove \Cref{thm:hashing-err}. 

\begin{proof}[Proof of \Cref{thm:hashing-err}]
By \Cref{lem:inverse-lipschitz}, we have
\begin{align*}
\E[\|\bn - \bhn\|_1] 
&~\leq~ 4 \cdot  \E[\|\Gamma^B(\bn) - \Gamma^B(\bhn)\|_1] \\
&~\leq~ 4 \cdot  \left(\E[\|\Gamma^B(\bn) - \bphi_{\geq}^{\bhn^{\red}}\|_1 + \|\bphi_{\geq}^{\bhn^{\red}} - \Gamma^B(\bhn)\|_1]\right) \\
&~\leq~ 8 \cdot  \left(\E[\|\Gamma^B(\bn) - \bphi_{\geq}^{\bhn^{\red}}\|_1]\right) \qquad \qquad \qquad (\text{By definition of } \bhn) \\
&~\leq~ 8 \cdot  \left(\E[\|\bphi_{\geq}^{\bn^{\red}} - \bphi_{\geq}^{\bhn^{\red}}\|_1 + \|\Gamma^B(\bn) - \bphi_{\geq}^{\bn^{\red}}\|_1 ]\right) \\
&~\leq~ O(\E[\|\bn^{\red} - \bhn^{\red}\|_1] + n \log n / \sqrt{B}) \quad (\text{From \Cref{lem:hash-concen}}). \qedhere
\end{align*}
\end{proof}

\section{Estimating Symmetric Properties of Discrete Distributions}\label{sec:ml-applications}

In this section we show how to use  \Cref{thm:hashing-err} for the task of estimating symmetric properties of discrete distributions over $[k]$. Here, a distribution property is {\em symmetric} if it remains unchanged under relabeling of the domain symbols. For example, a notable such property is the Shannon entropy of a distribution $\cD$ defined as $H(\cD) := \sum_x \cD(x) \log \frac{1}{\cD(x)}$, a central object in information theory, machine learning, and statistics. If the support is unbounded, estimating $H(\cD)$ is impossible with any finite number of samples. Our goal is to estimate the entropy of a distribution $\cD \in \Delta_k$ up to an additive $\pm \alpha$ error, where $\Delta_k$ denotes the set of all distributions over $[k]$.

One of the key ideas in the literature (e.g.,~\cite{AcharyaKSZ20}) is to design {\em low sensitivity} estimators $\hf : \cX^n \to \R$ for the desired symmetric distribution property $f : \Delta_k \to \R$.
The (non-private) \emph{sample complexity} of a property $f : \Delta_k \to \R$ using estimator $\hf$, denoted by $C_{\hf}(f, \alpha)$, is the smallest number of samples $n$ needed to estimate $f(\cD)$ upto accuracy $\alpha$ with probability at least $0.9$, that is, $\Pr[|\hf(S) - f(\cD)| > \alpha] < 0.1$.%
\footnote{The choice of $0.1$ is arbitrary; using the ``median trick'', we can boost the success probability to $1 - \beta$ with an additional multiplicative $\log(1/\beta)$ more samples.}
The \emph{sensitivity} of an estimator $\hf$ is $\Delta_{n,\hf}$, which is the smallest value for which it holds for adjacent datasets $S \sim S'$ each with $n$ elements, that $|\hf(S) - \hf(S')| \le \Delta_{n, \hf}$. Let $D_{\hf}(\alpha, \eps) := \min\{n \mid \Delta_{n, \hf} \le 0.1\alpha / E(n,\eps)\}$, for $E(n,\eps)$ defined in \Cref{cor:hashing-err}.

We will only consider {\em symmetric} estimators $\hf$, for which we will abuse notation to use $\hf(\bn)$ to denote $\hf(S)$ for any dataset $S$ that corresponds to the anonymized histogram $\bn$.

\begin{theorem}\label{thm:symm-prop-estimation}
For all $0 < \eps \leq O(1), \delta \in (0, 1]$, for any symmetric distribution property $f$, and any symmetric estimator $\hf$, there exists an $\eps$-DP mechanism in the pan-private model and $(\eps,\delta)$-DP mechanism in the shuffle DP model, such that $\Pr_{S \sim \cD^n}[|\cM(S) - f(\cD)| > \alpha] < 0.2$ with sample complexity 
\[
O\inparen{C_{\hf}\inparen{f, \frac\alpha2} + D_{\hf}\inparen{\frac\alpha2, \eps}}.
\]
\end{theorem}
\begin{proof}
Let $\bn$ denote the anonymized histogram corresponding to the sampled dataset $S$. The mechanism $\cM$ simply outputs $\hf(\bhn)$ for $\bhn$ returned by \Cref{alg:hash-L1-loss}. Clearly the mechanism $\cM$ is DP due to the post-processing property.

We have that for a suitable $n = O\inparen{C_{\hf}\inparen{f, \frac\alpha2} + D_{\hf}\inparen{\frac\alpha2, \eps}}$, it holds that
\begin{align*}
\Pr\insquare{\inabs{\hf(\bn) - f(\cD)} > \frac\alpha2} ~\le~ 0.1 \qquad \text{and} \qquad
\Pr\insquare{\inabs{\hf(\bhn) - \hf(\bn)} > \frac\alpha2 } ~\le~ 0.1,
\end{align*}
where the first inequality holds by definition of $C_{\hf}(f, \alpha/2)$, and the second inequality holds because $|\hf(\bhn) - \hf(\bn)| \le \Delta_{n,\hf} \cdot \|\bhn - \bn\|_1$ and by the guarantee of \Cref{thm:hashing-err} and Markov's inequality $\Pr[\|\bhn - \bn\|_1 > 10 E(n, \eps)] \le 0.1$.
By a union bound, we have  $\Pr[|\hf(\bhn) - \hf(\bn)| > \alpha] \le 0.44$. 
\end{proof}

We get the following sample complexity bounds for private estimation of Shannon entropy in the pan-private and shuffle DP models, as an immediate application of \Cref{thm:symm-prop-estimation}.

\begin{corollary}[Proof in \Cref{apx:ml-applications-proofs}]\label{cor:entropy-sample-complexity}
For all $0 < \eps \leq O(1), \delta \in (0, 1]$, there exists an $\eps$-DP mechanism in the pan-private model and $(\eps,\delta)$-DP mechanism in the shuffle DP model, that can estimate the entropy of $\cD \in \Delta_k$ up to an additive error of $\pm \alpha$ with a sample complexity of
\[
\textstyle\min\limits_{\lambda \in (0, \frac12)} \set{
O\inparen{\frac{k}{\alpha} + \frac{\log^2 k}{\alpha^2} + \frac{\log^{3.5}(1/(\alpha \eps))}{\alpha^2 \eps^5}}\ , \quad
O\inparen{\frac{k}{\lambda^2 \alpha \log k} + \frac{\log^2 k}{\alpha^2} + \inparen{\frac{\log^{1.5}(1/(\alpha \eps))}{\alpha^2 \eps^5}}^{1/(1-2\lambda)}}}\,.
\]
\end{corollary}
These bounds have the same dependence on $k$ as in the work of Acharya et al.~\cite{AcharyaKSZ20}. The dependence on $\alpha$ and $\eps$ is slightly worse due to a higher cost of sensitivity in our setting, and the worse dependence on $\eps$ in our guarantees in \Cref{cor:hashing-err}.

We also derive sample complexity bounds for private estimation of support coverage and support size. The \emph{support coverage} of a distribution $\cD$ and an integer $m$ is defined as $S_m(\cD) := \sum_{x \in \supp(\cD)}(1 - (1 - \cD(x))^m)$, i.e., the expected number of distinct elements seen if we draw $m$ i.i.d. samples from $\cD$. Here we would like to estimate $S_m(\cD)$ to within an additive error of $\pm \alpha m$. The {\em support size} of a discrete distribution $\cD$ is the number of atoms $x$ such that $\cD(x) > 0$. In general, this is impossible with finite sample since some atom of $\cD$ might have an arbritrarily small probability mass. To avoid this, we follow prior work and consider only probability distributions in $\Delta_{\geq 1/K} := \{\cD \mid \forall x \in \supp(\cD), \cD(x) \geq 1/K\}$, i.e., those with non-zero mass of at least $1/K$ at every atom, for some $K$. We defer the details of the exact sample complexity bounds and the proofs to \Cref{apx:ml-applications-proofs}.

\section{Conclusions and Future Directions}\label{sec:discussion}

In this paper, we give simple algorithms for privately computing anonymized histograms.  Our algorithms can be implemented in the shuffle and pan-private models. There are several immediate open questions as discussed below.

Our upper bounds have a dependency of $O(1/\eps^{2.5})$ in the $\ell_1$-error case and $O(1/\eps^{3.5})$ in the $\ell_2^2$-error case; it is unclear if these are tight. Similarly, there is a lower order multiplicative term of $O(\log n)$ and $O(\log^2 n)$ in our $\ell_1$- and $\ell_2^2$-error bounds respectively. Closing these gaps would be an interesting next step; these would also lead to improvements to the sample complexity bounds on private estimation of symmetric distribution properties, such as the Shannon entropy (\Cref{cor:entropy-sample-complexity}).

While our analysis of the large domain case relies on the hash function being uniformly random, it is conceivable that this is not required for the approach to work. It will be interesting if this approach, perhaps with some modifications, can also work with a weaker family of hash functions such as a pairwise-independent hash family.

Note also that our shuffle DP algorithm has $\delta > 0$, i.e., approximate-DP. This may not be necessary: there are a couple of recent algorithms for computing histogram with shuffle DP with $\delta = 0$~\cite{GhaziGKMPV20,CY21}. Our post-processing approach does not immediately apply to these algorithms because the noise to each count is not a discrete Laplace noise (and in fact is not even an independent additive noise). Adapting our approach to their setting is another interesting research direction.

\newpage

\bibliographystyle{abbrv}
\bibliography{main.bbl}
\newpage

\appendix

\section{Missing Proofs from \Cref{sec:large-domain}}

\subsection{Proof of \Cref{lem:inverse-lipschitz}}\label{apx:inverse-lipschitz-proof}

\inverselipschitz*

We start with the following observation that $\Gamma^B(\bn)_\ell := \E [\varphi^{\bn^{\red}}_{\ge \ell}]$ only depends on $\phi_{\geq 1}^{\bn}, \dots, \phi_{\geq \ell}^{\bn}$ . This is simply because changing an entry of value at least $\ell$ to some other value at least $\ell$ does not change whether, after hashing, the corresponding hash bucket exceeds $\ell$.
\begin{observation} \label{obs:dependent-cut}
$\Gamma^B(\bn)_\ell$ does \emph{not} depend on $\phi^{\bn}_{\geq r}$ for any $r > \ell$.
\end{observation}

Before can prove ~\Cref{lem:inverse-lipschitz}, we will also need to show the following lemma, which is even stronger than Lipschitzness.

\begin{lemma} \label{lem:lipschitz}
Let $\bbn, \tbn$ be anonymized histograms such that $\|\tbn - \bbn\|_1 = 1$, $\|\bbn\|_1 = n$, $\|\tbn\|_1 = n+1$, and $r \in [n]$ be such that $\phi_{\geq r}^{\tbn} = \phi_{\geq r}^{\bbn} + 1$. Then, we have
\begin{itemize}[leftmargin=6mm,nosep]
\item $\Gamma^B(\tbn)_\ell ~=~ \Gamma^B(\bbn)_\ell$\ for all $\ell < r$.
\item $\Gamma^B(\tbn)_r ~\geq~ \Gamma^B(\bbn)_r + 1 - n/B$.
\item $\sum_{\ell > r} |\Gamma^B(\tbn)_\ell - \Gamma^B(\bbn)_\ell| ~<~ n/B$.
\end{itemize}
Specifically, these also imply that $\|\Gamma^B(\tbn) - \Gamma^B(\bbn)\|_1 \leq 1$.
\end{lemma}

\begin{proof}
Let $\bbn^{\red}$ and $\tbn^{\red}$ be the random variables as defined at the beginning of \Cref{sec:large-domain}, but with starting anonymized histograms $\bbn, \tbn$ respectively. Furthermore, let $j$ be the entry such that $\tn^{(j)} = \ban^{(j)} + 1$.

Recall that $H$ is our random hash function. Consider $\bbn^{\red}(H)$ and $\tbn^{\red}(H)$ resulting from the same hash function $H$. By definition, we have $\Gamma^B(\tbn) - \Gamma^B(\bbn) = \E_H[\bphi^{\tbn^{\red}(H)}_{\geq} - \bphi^{\bbn^{\red}(H)}_{\geq}]$.

Furthermore, we have  $\bphi^{\tbn^{\red}(H)}_{\geq} - \bphi^{\bbn^{\red}(H)}_{\geq} = \ind_{\tn_{H(j)}^{\red}}$, where $\tn_{H(j)}^{\red}$ is the value of the bucket $H(j)$ to which $j$ gets hashed. Note that with probability at least $1 - n/B$, there is no collision with $j$ and therefore $\tn_{H(j)}^{\red} = r$. Otherwise, with probability at most $n/B$, there is a collision and $\tn_{H(j)}^{\red} > r$. This implies the desired bounds.
\end{proof}

We are now ready to prove \Cref{lem:inverse-lipschitz}.

\begin{proof}[Proof of \Cref{lem:inverse-lipschitz}]
Let $\bn^{\min}$ be such that $\phi_{\geq r}^{\bn^{\min}} = \min\{\phi_{\geq r}^{\bn}, \phi_{\geq r}^{\bn'}\}$ for all $r \in \N$. Note that $\|\bn^{\min} - \bn\|_1 + \|\bn^{\min} - \bn'\|_1 = \|\bn - \bn'\|_1$. Assume w.l.o.g. that $\|\bn^{\min} - \bn\|_1 \geq \|\bn^{\min} - \bn'\|_1$. The previous inequality implies that $\|\bn^{\min} - \bn\|_1 \geq \|\bn - \bn'\|_1 / 2$.

Now, let $S := \{r \mid \phi^{\bn}_{\geq r} > \phi^{\bn^{\min}}_{\geq r}\}$. Consider the following hybrids for $j = 0, \dots, n$: let $\bn^j$ be such that $\bphi^{\bn^j}_{\geq} = (\phi^{\bn}_{\geq 1}, \dots, \phi^{\bn}_{\geq j}, \phi^{\bn^{\min}}_{\geq j + 1}, \dots, \phi^{\bn^{\min}}_{\geq n})$. By \Cref{lem:lipschitz} and the definition of $S$, we have
\begin{align*}
\sum_{r \in S} \left(\Gamma^B(\bn^{j+1})_r - \Gamma^B(\bn^j)_r\right) \geq \frac{3}{4} \left(\phi^{\bn^{j+1}}_{\geq j + 1} - \phi^{\bn^j}_{\geq j + 1} \right),
\end{align*}
since $\|\bn^{\min}\|_1 \le \|\bn\|_1 \le n$. By summing the above over all $j = 0, \dots, n - 1$, we have
\begin{align} \label{eq:hybrid-first}
\sum_{r \in S} \left(\Gamma^B(\bn)_r - \Gamma^B(\bn^{\min})_r\right) \geq \frac{3}{4} \|\bn^{\min} - \bn\|_1.
\end{align}

Similarly, we can consider the following hybrids for $j = 0, \dots, n$: let $\bn'^j$ be such that $\bphi^{\bn'^j}_{\geq} = (\phi^{\bn'}_{\geq 1}, \dots, \phi^{\bn'}_{\geq j}, \phi^{\bn^{\min}}_{\geq j + 1}, \dots, \phi^{\bn^{\min}}_{\geq n})$. By \Cref{lem:lipschitz} and the definition of $S$ (which implies that $\phi^{\bn'}_{\geq r} = \phi^{\bn^{\min}}_{\geq r}$ for all $r \in S$), we have
\begin{align*}
\sum_{r \in S} \left(\Gamma^B(\bn'^{j+1})_r - \Gamma^B(\bn'^j)_r\right) \leq \frac{1}{4} \left(\phi^{\bn'^{j+1}}_{\geq j + 1} - \phi^{\bn'^j}_{\geq j + 1} \right).
\end{align*}
Again, by summing the above over all $j = 0, \dots, n - 1$, we have
\begin{align} \label{eq:hybrid-second}
\sum_{r \in S} \left(\Gamma^B(\bn')_r - \Gamma^B(\bn^{\min})_r\right) \leq \frac{1}{4} \|\bn^{\min} - \bn'\|_1.
\end{align}

Subtracting~\eqref{eq:hybrid-second} from~\eqref{eq:hybrid-first}, we complete the proof as follows:
\begin{align*}
\|\Gamma^B(\bn) - \Gamma^B(\bn')\|_1 &~\geq~ \sum_{r \in S} \Gamma^B(\bn)_r - \Gamma^B(\bn')_r
\enspace~\geq~\enspace \frac{3}{4} \|\bn^{\min} - \bn\|_1 - \frac{1}{4} \|\bn^{\min} - \bn'\|_1 \\
&~\geq~ \frac{1}{2} \|\bn^{\min} - \bn\|_1 \enspace~\geq~\enspace \frac{1}{4} \|\bn - \bn'\|_1.
\qedhere
\end{align*}
\end{proof}

\subsection{Proof of \Cref{lem:var-phi-hashed-term}}\label{apx:var-phi-hashed-term-proof}

\varphihashedterm*

\begin{proof}[Proof of \Cref{lem:var-phi-hashed-term}]
Note that $\phi_{\geq r}^{\bn^{\red}}$ is the number of hash buckets whose total value is at least $r$. Let $X$ denote the number of hash buckets whose maximum value hashed to it is at least $r$, i.e., $X := |\{i \in [B] \mid \max_{j \in H^{-1}(i)} n^{(j)} \geq r\}|$. Furthermore, for $t \in [r - 1]$, let $Y_t$ denote the number of buckets whose total value is at least $r$ and the maximum value hashed to it is equal to $t$, i.e., $Y_t := |\{i \in [B] \mid \max_{j \in H^{-1}(i)} n^{(j)} = t, \sum_{j \in H^{-1}(i)} n^{(j)} \geq r\}|$.

It is easy to see that $X, Y_1, \dots, Y_{r - 1}$ are negatively correlated. Therefore,
\begin{align*}
\Var[\phi_{\geq r}^{\bn^{\red}}] ~=~ \Var[X + Y_1 + \cdots + Y_{r - 1}] ~\leq~ \Var[X] + \sum_{t \in [r - 1]} \Var[Y_t].
\end{align*}

We will next bound each term in the RHS above.

\paragraph{\boldmath Bounding $\Var[X]$.} To bound $\Var[X]$, further note that $X = \phi^{\bn}_{\geq r} - \sum_{i \in [B]} Z_i$ where $Z_i := \max\{0, |\{j \in H^{-1}(i) \mid n^{(j)} \geq r\}| - 1\}$. It is again easy to see that $Z_i$'s are negatively correlated. Therefore,
\begin{align*}
\Var[X] = \Var\left[\sum_{i \in [B]} Z_B\right] \leq \sum_{i \in [B]} \Var[Z_i] \leq \sum_{i \in [B]} \E[Z_i^2].
\end{align*}
It is also easy to verify that, for any $k \in \N$, $\Pr[Z_i = k] = \binom{\phi^{\bn}_{\geq r}}{k+1}(1/B)^{k+1}(1 - 1/B)^{\phi^{\bn}_{\geq r} - k - 1} \leq (\phi^{\bn}_{\geq r} / B)^{k + 1} / (k + 1)!$. Therefore, we have
\begin{align*}
\E[Z_i^2] 
&\leq \sum_{k \in \N} k^2 \cdot (\phi^{\bn}_{\geq r} / B)^{k + 1} / (k + 1)! 
\enspace~\leq~\enspace 2 \cdot \sum_{k \in \N} (\phi^{\bn}_{\geq r}/B)^{k+1} \\
&\leq 4 \inparen{ \frac{\phi^{\bn}_{\geq r}}{B}}^2 
\enspace~\leq~\enspace 4 \inparen{\frac{n}{Br}}^2 
\enspace~=~\enspace \frac{4}{r^2} \cdot \left(\frac{n}{B}\right)^2.
\end{align*}
Plugging this back into the previous inequality, we have
\begin{align*}
\Var[X] ~\leq~ \frac{4}{r^2} \cdot \frac{n^2}{B}.
\end{align*}

\paragraph{\boldmath Bounding $\Var[Y_t]$.} We may write $Y_t$ as $\sum_{i \in [B]} W_i$ where $W_i := \ind[\max_{j \in H^{-1}(i)} n^{(j)} = t, \sum_{j \in H^{-1}(i)} n^{(j)} \geq r]$. Again, the $W_i$'s are negatively correlated. Therefore, we have
\begin{align}
\Var[Y_t] = \Var \left[\sum_{i \in [B]} W_i \right] \leq \sum_{i \in [B]} \Var[W_i] \leq  \sum_{i \in [B]} \Pr[W_i = 1], \label{eq:var-yt-expanded}
\end{align}
where the last inequality uses the fact that $W_i$ is a Bernoulli r.v.

For each $t$, let $S_t \subseteq \N$ denote the set of indices $j$ for which $n^{(j)} = t$. Furthermore, let $S_{\leq t} := S_1 \cup \cdots \cup S_t$.
Notice that 
\begin{align}
\Pr[W_i = 1] 
&~=~ \Pr\left[\max_{j \in H^{-1}(i)} n^{(j)} = t \text{ and } \sum_{j \in H^{-1}(i)} n^{(j)} \geq r\right] \nonumber \\
&~=~ \Pr\left[\max_{j \in H^{-1}(i)} n^{(j)} = t \text{ and } \sum_{j \in H^{-1}(i) \cap S_{\leq t}} n^{(j)} \geq r\right] \nonumber \\
&~=~ \Pr\left[\exists j^* \in S_t \text{ s.t } H(j^*) = i \text{ and } \sum_{j \in H^{-1}(i) \cap S_{\leq t}} n^{(j)} \geq r\right] \nonumber \\
(\text{Union bound}) &~\leq~ \sum_{j^* \in S_t} \Pr\left[H(j^*) = i \text{ and } \sum_{j \in H^{-1}(i) \cap S_{\leq t}} n^{(j)} \geq r\right] \nonumber \\
&~=~ \sum_{j^* \in S_t} \Pr\left[H(j^*) = i \text{ and } \sum_{j \in H^{-1}(i) \cap (S_{\leq t} \setminus \{j^*\})} n^{(j)} \geq r - t\right] \nonumber \\
(\text{Independence of hash values}) &~=~ \sum_{j^* \in S_t} \Pr[H(j^*) = i] \Pr\left[\sum_{j \in H^{-1}(i) \cap (S_{\leq t} \setminus \{j^*\})} n^{(j)} \geq r - t\right] \nonumber \\
&~=~ \sum_{j^* \in S_t} \frac{1}{B} \cdot \Pr\left[\sum_{j \in H^{-1}(i) \cap (S_{\leq t} \setminus \{j^*\})} n^{(j)} \geq r - t\right] \nonumber \\
&~\leq~ \sum_{j^* \in S_t} \frac{1}{B} \cdot \Pr\left[\sum_{j \in H^{-1}(i) \cap S_{\leq t}} n^{(j)} \geq r - t\right] \nonumber \\
&~=~ \frac{\phi^{\bn}_t}{B} \cdot \Pr\left[\sum_{j \in H^{-1}(i) \cap S_{\leq t}} n^{(j)} \geq r - t\right] \nonumber \\
&~=~ \frac{\phi^{\bn}_t}{B} \cdot \Pr\left[\sum_{j \in S_{\leq t}} n^{(j)} \cdot U_j \geq r - t\right], \label{eq:var-wi-expanded}
\end{align}
where $U_j$ denote the indicator $\ind[H(j) = i]$.

We will next bound the last term based on two cases:
\begin{itemize}[leftmargin=4mm,nosep]
\item Case I: $t \geq r/2$. Note that
\begin{align*}
\E\left[\sum_{j \in S_{\leq t}} n^{(j)} \cdot U_j\right] = \left(\sum_{j \in S_{\leq t}} n^{(j)}\right) \cdot \frac{1}{B} \leq \frac{n}{B}.
\end{align*}
Therefore, we may apply Markov's inequality to get
\begin{align*}
\Pr\left[\sum_{j \in S_{\leq t}} n^{(j)} \cdot U_j \geq r - t\right] \leq \frac{n}{B(r - t)} \leq \frac{2n}{B} \cdot \frac{t}{r(r - t)},
\end{align*}
where the latter inequality follows from $t \geq r/2$.
\item Case II: $t < r/2$. Since $U_j$'s are independent, we can also compute the variance of the sum as
\begin{align*}
\Var\left[\sum_{j \in S_{\leq t}} n^{(j)} \cdot U_j\right]
&= \sum_{j \in S_{\leq t}} \Var[n^{(j)} \cdot U_j] 
\enspace~\leq~\enspace \sum_{j \in S_{\leq t}} \frac{(n^{(j)})^2}{B} \\
&= \frac{1}{B} \sum_{\ell \in [t]} \ell^2 \cdot \phi^{\bn}_\ell 
\enspace~\leq~\enspace \frac{t}{B} \sum_{\ell \in [t]} \ell \cdot \phi^{\bn}_\ell 
\enspace~\leq~\enspace \frac{tn}{B}.
\end{align*}
Note also that we have $r - t - n/B > (r - t)/2 \geq r/4$, where we used $t < r/2$.  
We may now apply Chebyshev's inequality to get
\begin{align*}
\Pr\left[\sum_{j \in S_{\leq t}} n^{(j)} \cdot U_j \geq r - t\right] \leq \frac{tn/B}{(r/4)^2} \leq \frac{16n}{B} \cdot \frac{t}{r(r -t)}.
\end{align*}
\end{itemize}
Therefore, in both cases, we have $\Pr\left[\sum_{j \in S_{\leq t}} n^{(j)} \cdot U_j \geq r - t\right] \leq \frac{16n}{B} \cdot \frac{t}{n - t}$. Combining this with~\eqref{eq:var-yt-expanded} and~\eqref{eq:var-wi-expanded}, we get 
\begin{align*}
\Var[Y_t] \leq B \cdot \frac{\phi^{\bn}_t}{B} \cdot \frac{16n}{B} \cdot \frac{t}{r(r -t)} = \frac{16n}{B} \cdot \frac{t \cdot \phi^{\bn}_t}{r(r - t)}.
\end{align*}
By summing up our bounds on $X$ and $Y_t$'s, we arrive at the desired bound
\begin{align*}
\Var[\phi_{\geq r}^{\bn^{\red}}] 
& \leq \frac{16n}{B} \cdot \left(\frac{n}{r^2} + \sum_{t \in [r - 1]} \frac{t \cdot \phi^{\bn}_t}{r(r - t)}\right).
\qedhere
\end{align*}
\end{proof}

\section{\boldmath Extension to $\ell_2^2$-error}\label{apx:l22_error}

A different error measure used in~\cite{HayRMS10} is the $\ell_2^2$-error, defined as $\|\bn - \bhn\|_2^2 = \sum_{j \in [D]} (n^{(j)} - \hn^{(j)})^2$. In this section, we show the flexibility of our method by showing that we can get an error of $\tO_\eps(\sqrt{n})$, nearly matching that of~\cite{HayRMS10}.

\subsection{Small Domain Case}

We start with the setting $D \leq \tO(n)$ which does not require hashing.

The idea is to constrain our estimate anonymized histogram such that the $\ell_\infty$-distance from the discrete Laplace-noised histogram is not too large.  A full description is presented in \Cref{alg:L22-loss}.

\begin{algorithm}[h]
\caption{Anonymized Histogram Estimator w.r.t. $\ell^2_2$ loss.}
\label{alg:L22-loss}
\begin{algorithmic}
\STATE {\bf Input:} Discrete Laplace-noised histogram $\bh'$, i.e., $h'_j \sim h_j + \DLap(p)$\\
{\bf Parameter:} $\gamma = 10 \log(2nD) / \log(1/p)$\\[2mm]
\FOR{$r \in [n]$}
    \STATE $\hphi_{\geq r} \gets \sum_{j \in [D]} f(h'_j - r)$, where $f$ is defined in \eqref{eq:post-process-f}
\ENDFOR
\RETURN $\bhn$ that minimizes $\|\bphi^{\bhn}_{\geq} - \hbphi_{\geq}\|_1$ subject to $\|\bn_{\bh'} - \bhn\|_\infty \le \gamma$ and $\|\bhn\|_1 = n$. (If no such $\bhn$ exists, output the all-zeros histogram.)
\end{algorithmic}
\end{algorithm}

\begin{theorem} \label{thm:err-l22}
Let $\bhn$ be the output of \Cref{alg:L22-loss}. We have $$\E[\|\bhn - \bn \|_2^2] \leq O\left(\frac{\log(nD)}{\log(1/p)} \cdot \sqrt{C_p(n+D)\log n}\right) + O(1).$$
\end{theorem}

Plugging in $p = e^{-\eps/2}$ (sufficient for $\eps$-DP), we have the bound of $O(\sqrt{(n + D)\log n} \cdot \log(nD) / \eps^{3.5})$ for $\eps \leq 1$.

\begin{proof}
First, the standard concentration of the noise implies that
\begin{align*}
\Pr[\|\bn - \bn_{\bh'}\|_\infty > \gamma] \leq 0.1/n^2.
\end{align*}
We thus have
\begin{align*}
\E[\|\bhn - \bn \|_2^2] &\leq \Pr[\|\bn - \bn_{\bh'}\|_\infty \leq \gamma] \cdot \E[\|\bhn - \bn \|_2^2 \mid \|\bn - \bn_{\bh'}\|_\infty \leq \gamma] + \Pr[\|\bn - \bn_{\bh'}\|_\infty > \gamma] \cdot n^2 \\
&\leq \Pr[\|\bn - \bn_{\bh'}\|_\infty \leq \gamma] \cdot \E[\|\bhn - \bn \|_2^2 \mid \|\bn - \bn_{\bh'}\|_\infty \leq \gamma] + O(1).
\end{align*}
Now, recall that $\|\bn - \bhn\|_2^2 \leq \|\bn - \bhn\|_\infty \cdot \|\bn - \bhn\|_1$. Plugging this into the above, we get
\begin{align*}
\E[\|\bhn - \bn \|_2^2] 
&\leq \Pr[\|\bn - \bn_{\bh'}\|_\infty \leq \gamma] \cdot \E[\|\bhn - \bn \|_1 \cdot \|\bhn - \bn\|_\infty \mid \|\bn - \bn_{\bh'}\|_\infty \leq \gamma] + O(1) \\
&\leq \Pr[\|\bn - \bn_{\bh'}\|_\infty \leq \gamma] \cdot \E[\|\bhn - \bn \|_1 \cdot 2\gamma \mid \|\bn - \bn_{\bh'}\|_\infty \leq \gamma] + O(1) \\
&\leq 2\gamma \cdot \Pr[\|\bn - \bn_{\bh'}\|_\infty \leq \gamma] \cdot \E[2 \|\bphi^{\bn}_{\geq} - \hbphi_{\geq}\|_1 \mid \|\bn - \bn_{\bh'}\|_\infty \leq \gamma] + O(1) \\
&\leq 4\gamma \cdot \E[\|\bphi^{\bn}_{\geq} - \hbphi_{\geq}\|_1] + O(1) \\
&\leq 4\gamma \cdot O\left(\sqrt{C_p(n+D)\log n}\right) + O(1), \\
&= O\left(\frac{\log(nD)}{\log(1/p)} \cdot \sqrt{C_p(n+D)\log n}\right) + O(1),
\end{align*}
where the second inequality follows from the constraint on $\bhn$ and the last inequality follows from our analysis in the $\ell_1$-error case (\Cref{thm:err-final}).
\end{proof}

\subsection{Large Domain Case}

We next move on to the case $D \gg n$, which will require random hashing.

\subsubsection{\boldmath Barrier to Extending to the $\ell_2^2$-error}

It turns out that, unlike the $\ell_1$-error case, using a single noisy hashed histogram with $B = \tO(n)$ is not sufficient for us to get $\tO(\sqrt{n})$ $\ell_2^2$-error. Before we provide the fix for this, let us briefly sketch why this is the case.

Let $c = \lceil 10 \sqrt{B} \rceil$ and $q = \lfloor n/c \rfloor$. Consider the following two datasets (before hashing):
\begin{itemize}[leftmargin=4mm,nosep]
\item There are $c$ items, each with value $q$.
\item There are $c - 2$ items each with value $q$ and one additional item with value $2q$.
\end{itemize}
It is not hard to see (from birthday paradox) that, after randomly hashing into $B$ buckets, it is impossible to distinguish the two cases with advantage more than 0.1 (even without any discrete Laplace noise). This implies that the expected  $\ell_2^2$-error must be at least $\Omega(q^2) = \Omega(n^2 / B)$. For $B = \tO(n)$, this is at least $\tOmega(n)$.

\subsubsection{Two-Hash Approach}

We now sketch an approach for large $D$. Here we would use two hashes. The first is with $B_1 = \tO(n)$ buckets similar to before and but the second with a much larger, say, $B_2 = O(n^4)$ buckets. We then use the $\ell_1$ approach on the first hash with the additional $\ell_\infty$-constraint on the second hash. The main point here is that w.h.p. there would be no collision at all in the second hash; therefore, the $\ell_\infty$-constraint will be valid. A similar analysis to the small domain case shows that this only adds $O_\eps(\log n)$ multiplicative overhead on the expected $\ell_2^2$-error (compared to the $\ell_1$-error).

More precisely, our approach is presented in \Cref{alg:hash-L22-loss}.

\begin{algorithm}[h]
\caption{Anonymized Histogram Estimator w.r.t. $\ell_2^2$ loss, for large domains.}
\label{alg:hash-L22-loss}
\begin{algorithmic}
\STATE {\bf Input:} Two discrete Laplace-noised histograms $\tbh^{1, \red}$ and $\tbh^{2, \red}$ given as
\begin{itemize}[nosep]
	\item $\tildeh^{1, \red}_j \sim h^{1, \red}_j + \DLap(p)$ where the random hash has $B_1$ buckets, and
	\item $\tildeh^{2, \red}_j \sim h^{2, \red}_j + \DLap(p)$ where the random hash has $B_2$ buckets.
\end{itemize}
{\bf Parameter:} $\gamma = 10 \log(2n B_2) / \log(1/p)$\\[2mm]

\STATE Compute an estimate $\bhn^1$ by running \Cref{alg:hash-L1-loss} on $\tbh^1$.
\RETURN $\bhn$ that minimizes $\|\bhn - \bhn^{1}\|_1$ subject to $\|\bn_{\tbh^{2,\red}} - \bhn\|_\infty \le \gamma$ and $\|\bhn\|_1 = n$. (If no such $\bhn$ exists, just output the all-zero histogram.)
\end{algorithmic}
\end{algorithm}

\begin{theorem} \label{thm:hashing-l22err}
Let $\bhn$ be the output of \Cref{alg:hash-L22-loss}. We have $$\E[\|\bhn - \bn \|_2^2] \leq O\left(\frac{\log(nB_2)}{\log(1/p)} \cdot \inparen{\sqrt{C_p (n+B_1) \log n} + \frac{n \log n}{\sqrt{B_1}}}\right) + O\inparen{\frac{n^4}{B_2} + 1}.$$
\end{theorem}

Plugging in $p = e^{-\eps/4}$ (sufficient for $\eps$-DP), $B_1 = n \sqrt{\log n}$ and $B_2 = n^4$, we have the bound of $O(\sqrt{n} (\log n)^{7/4} / \eps^{3.5})$ for $\eps \leq 1$.

\begin{proof}
The probability that there is any collision in the second hash is at most $n^2/B_2$. Furthermore, standard concentration inequality of the noise implies that the noise added to any of $h^{2, \red}_j$ is greater than $\gamma$ is at most $0.1/n^2$. Therefore, by a union bound, we have
\begin{align*}
\Pr[\|\bn - \bn_{\tbh^{2, \red}}\|_\infty > \gamma] \leq n^2/B_2 + 0.1/n^2.
\end{align*}
The remaining analysis is now similar to that of~\Cref{thm:err-l22}. Specifically, we have
\begin{align*}
&\E[\|\bhn - \bn \|_2^2] \\
&\leq \Pr[\|\bn - \bn_{\tbh^{2, \red}}\|_\infty \leq \gamma] \cdot \E[\|\bhn - \bn \|_1 \cdot \|\bhn - \bn\|_\infty \mid \|\bn - \bn_{\tbh^{2, \red}}\|_\infty \leq \gamma] \\
&\qquad + \Pr[\|\bn - \bn_{\tbh^{2, \red}}\|_\infty > \gamma] \cdot n^2 \\
&\leq \Pr[\|\bn - \bn_{\tbh^{2, \red}}\|_\infty \leq \gamma] \cdot \E[2\|\bhn^1 - \bn \|_1 \cdot (2\gamma) \mid \|\bn - \bn_{\tbh^{2, \red}}\|_\infty \leq \gamma] + n^4/B_2 + 0.1 \\
&\leq (4\gamma) \cdot \E[\|\bhn^1 - \bn \|_1] + n^4/B_2 + 0.1 \\
&\leq O\inparen{\gamma\cdot \inparen{\sqrt{C_p (n+B_1) \log n} + \sqrt{n^2/B_1} \log n} + n^4/B_2 + 1}, 
\end{align*}
where the last inequality follows from \Cref{thm:hashing-err}.
\end{proof}

\section{Missing Proof from \Cref{sec:ml-applications}}\label{apx:ml-applications-proofs}

In this section, we prove sample complexity bounds for estimating symmetric distribution properties. In addition to entropy (\Cref{cor:entropy-sample-complexity}), we also show sample complexity bounds for support coverage (\Cref{cor:support-coverage-sample-complexity}) and support size (\Cref{cor:support-size-sample-complexity}).

\subsection{Entropy}

\begin{proof}[Proof of \Cref{thm:symm-prop-estimation}]
Acharya et al.~\cite{AcharyaKSZ20} consider two estimators for entropy estimation.
\begin{itemize}[leftmargin=4mm,nosep]
    \item The first estimator $\hat{H}$ they study is the entropy of the empirical distribution. This has a non-private sample complexity of $C_{\hat{H}}(H, \alpha) = O\inparen{\frac{k}{\alpha} + \frac{\log^2 k}{\alpha^2}}$. The sensitivity of this estimator is $O(\log n / n)$. Thus, we have $D_{\hat{H}}(\alpha, \eps)$ is the smallest $n$ that satisfies
    \[ \frac{\log n}{n} \cdot \frac{\sqrt{n} (\log n)^{3/4}}{\eps^{2.5}} \lesssim \alpha. \]
    And so, $D_{\hat{H}}(\alpha, \eps) \le O(\frac{\log^{3.5}(1/(\alpha \eps))}{\alpha^2 \eps^5})$. Thus, plugging this into \Cref{thm:symm-prop-estimation} yields a sample complexity of $O\inparen{\frac{k}{\alpha} + \frac{\log^2 k}{\alpha^2} + \frac{\log^{3.5}(1/(\alpha\eps))}{\alpha^2 \eps^5}}$.
    
    \item The second estimator $\hat{H}$ they study is based on the prior work of \cite{AcharyaDOS17}, which for any $\lambda \in (0, 1)$ has a non-private sample complexity of $C_{\hat{H}}(H, \alpha) = O\inparen{\frac{k}{\lambda^2 \alpha \log k} + \frac{\log^2 k}{\alpha^2}}$. The sensitivity of this estimator is $1/n^{1-\lambda}$. Thus, we have $D_{\hat{H}}(\alpha, \eps)$ is the smallest $n$ that satisfies
    \[ \frac{1}{n^{1-\lambda}} \cdot \frac{\sqrt{n} (\log n)^{3/4}}{\eps^{2.5}} \lesssim \alpha. \]
    And so, $D_{\hat{H}}(\alpha,\eps) \le O\inparen{\inparen{\frac{\log^{1.5}(1/(\alpha \eps)}{\alpha^2 \eps^5}}^{1/(1-2\lambda)}}$.
    Thus, plugging this into \Cref{thm:symm-prop-estimation} yields a sample complexity of $O\inparen{\frac{k}{\lambda^2 \alpha \log k} + \frac{\log^2 k}{\alpha^2} + \inparen{\frac{\log^{1.5}(1/(\alpha \eps)}{\alpha^2 \eps^5}}^{1/(1-2\lambda)}}$.\qedhere
\end{itemize}
\end{proof}

\subsection{Support Coverage}

The \emph{support coverage} of a distribution $\cD$ and an integer $m$ is defined as $S_m(\cD) := \sum_{x \in \supp(\cD)}(1 - (1 - \cD(x))^m)$, i.e., the expected number of distinct elements seen if we draw $m$ i.i.d. samples from $\cD$. Here we would like to estimate $S_m(\cD)$ to within an additive error of $\pm \alpha m$.

Our bounds are stated below. Note that, in the case of large $m$, our sample complexity is asymptotically the same as that of \cite{AcharyaKSZ20} but our dependency on $\alpha, \eps$ is worse in the case of smaller $m$.

\begin{corollary}[Support Coverage]\label{cor:support-coverage-sample-complexity}
For all $0 < \eps \leq O(1), \delta \in (0, 1]$, there exists an $(\eps,\delta)$-DP mechanism, in the pan-private and shuffle DP models, that can estimate $S_m(\cD)$ up to an additive error of $\pm \alpha m$ with a sample complexity of
\[
\begin{cases}
O\left(\frac{m\log(1/\alpha)}{\log(\alpha m \eps)}\right) & \text{if } m \geq \Omega\inparen{\frac{\log^{1.5}(1/(\alpha\eps))}{\alpha^2\eps^5}},\\
O\inparen{\frac{\log^{1.5}(1/(\alpha\eps))}{\alpha^2\eps^5}} & \text{if } m \leq O\inparen{\frac{\log^{1.5}(1/(\alpha\eps))}{\alpha^2\eps^5}}.
\end{cases}
\]
\end{corollary}

\begin{proof}
Throughout the proof, it will be simpler to think of estimating $\frac{S_m(\cD)}{m}$ to within an additive error of $\pm \alpha$.
Acharya et al.~\cite{AcharyaKSZ20} consider two estimators for support coverage estimation.
\begin{itemize}[leftmargin=4mm,nosep]
    \item {\bf Sparse Case.} Assuming that $m \geq \Omega\inparen{\frac{\log^2(1/(\alpha\eps))}{\alpha^2\eps^5}}$. When $m \geq 2n$, Acharya et al.~\cite{AcharyaKSZ20} (based on an earlier work \cite{Orlitsky-pnas}) gave an estimator $\hat{S}_m$, parameterized by $r = \log(3/\alpha)$ and $t = m/n - 1$, that has a non-private sample complexity of $C_{\hat{S}_n}(S_m, \alpha) = O\inparen{\frac{m \log(1/\alpha)}{\log m}}$. The sensitivity of this estimator is $\frac{1 + e^{r(t - 1)}}{m} = \frac{1 + e^{\log(3/\alpha)(m/n - 2)}}{m}$. Thus, we have $D_{\hat{S}_m}(\alpha, \eps)$ is the smallest $n$ that satisfies
    \[ \frac{1 + e^{\log(3/\alpha)(m/n - 2)}}{m} \cdot \frac{\sqrt{n} (\log n)^{3/4}}{\eps^{2.5}} \le \alpha\]
    And so, $D_{\hat{S}_m}(\alpha, \eps) \le O\left(\frac{m\log(1/\alpha)}{\log(\alpha m \eps)}\right)$. Thus, plugging this into \Cref{thm:symm-prop-estimation} gives rise to a sample complexity of $O\left(\frac{m\log(1/\alpha)}{\log(\alpha m \eps)}\right)$.
    
    \item {\bf Dense Case.} The second estimator $\hat{S}_m$ they study works when $n$ is a multiple of $m$. We remark that, although the estimator as described in~\cite{AcharyaKSZ20} does not seem to fit our setting, it can be viewed in our framework as follows. The examples are divided in to $n/m$ batches each of size $m$. We then build a histogram $\bh$ on $\supp(\cD) \times [n/m]$ where each example in batch $i$ appends $i$ to its original item. The final estimate is then $\phi^{\bh}_{\geq 1} / n$. It was shown in~\cite{AcharyaKSZ20} that the non-private sample complexity is $C_{\hat{S}_m}(S_m, \alpha) = O\inparen{1/\alpha^2}$.
    
    The sensitivity of this estimator is $1/n$. Thus, we have $D_{\hat{S}_m}(\alpha, \eps)$ is the smallest $n$ that satisfies
    \[ \frac{1}{n} \cdot \frac{\sqrt{n} (\log n)^{3/4}}{\eps^{2.5}} \le \alpha\]
    And so, $D_{\hat{S}_m}(\alpha,\eps) \le O\inparen{\frac{\log^{1.5}(1/(\alpha\eps))}{\alpha^2\eps^5}}$.
    Thus, plugging this into \Cref{thm:symm-prop-estimation} gives rise to a sample complexity of $O\inparen{\max\left\{m, \frac{\log^{1.5}(1/(\alpha\eps))}{\alpha^2\eps^5}\right\}}$, where the $m$ part comes from our constraint that $n \geq m$.\qedhere
\end{itemize}
\end{proof}

\subsection{Support Size}

Finally, we consider the problem of estimating the support size of $\cD$. In general, this is impossible with finite sample since some atom of $\cD$ might have an arbritrarily small probability mass. To avoid this, we follow prior work and consider only probability distributions in $\Delta_{\geq 1/K} := \{\cD \mid \forall x \in \supp(\cD), \cD(x) \geq 1/K\}$, i.e., those with non-zero mass of at least $1/K$ at every atom, for some $K$.  

Orlitsky et al.~\cite{Orlitsky-pnas} proved that the support coverage with $m \geq \Omega(K \cdot \log(3/\alpha))$ is a good estimate to the support size of any $\cD \in \Delta_{\geq 1/K}$ to within an error of $\pm \alpha K$. In particular, they showed that if $m \ge K \log(3/\alpha)$, then for any $\cD \in \Delta_{1/K}$, it holds that $|S_m(\cD) - S(\cD)| \le \alpha K/3$.
Combining this with \Cref{cor:support-size-sample-complexity}, we immediately get the following bounds for the latter.

\begin{corollary}[Support Size]\label{cor:support-size-sample-complexity}
For all $0 < \eps \leq O(1), \delta \in (0, 1]$, there exists an $(\eps,\delta)$-DP mechanism, in the pan-private and shuffle DP settings, that can estimate the support size of $\cD \in \Delta_{\geq 1/K}$ up to an additive error of $\pm \alpha K$ with a sample complexity of
\[
\begin{cases}
O\left(\frac{K \log^2(1/\alpha)}{\log(\alpha K \eps)}\right) & \text{if } K \geq \Omega\inparen{\frac{\log^{1.5}(1/(\alpha\eps))}{\alpha^2\eps^5 \log(1/\alpha)}},\\
O\inparen{\frac{\log^{1.5}(1/(\alpha\eps))}{\alpha^2\eps^5}} & \text{if } K \leq O\inparen{\frac{\log^{1.5}(1/(\alpha\eps))}{\alpha^2\eps^5\log(1/\alpha)}}.
\end{cases}
\]
\end{corollary}

\section{Fast Post-Processing Algorithms}

So far we have not focused on the running time of our post-processing algorithms. Nonetheless, it is not hard to see that our algorithms run in polynomial time. Below, we show that our algorithms can even be made to run in $\tO_{\eps}(D + n)$-time for the $\ell_1$-error case.

\subsection{Small Domain Case}\label{apx:L1-loss-fast}

We start with how to implement \Cref{alg:L1-loss} in $O(D + n \log n)$ time. The first observation is that the expression for each $\hphi_{\geq r}$ can be easily computed in constant time if we precompute the number of buckets above a certain threshold beforehand. The second observation is that the last step is exactly the same as the so-called \emph{$\ell_1$-isotonic regression} problem (by viewing the cumulative prevalence $\phi_{\geq 1}, \dots, \phi_{\geq n}$ as the variables), for which an $O(n \log n)$-algorithm is known \cite{rote19isotonic}.  A full description is given in \Cref{alg:L1-loss-fast}.

\newcommand{\myComment}[1]{\textcolor{black!60}{\{#1\}}}
\begin{algorithm}[h]
\caption{Efficient Anonymized Histogram Estimator w.r.t. $\ell_1$ loss.}
\label{alg:L1-loss-fast}
\begin{algorithmic}
\STATE {\bf Input:} Discrete Laplace-noised histogram $\bh'$, i.e., $h'_j \sim h_j + \DLap(p)$\\[2mm]
\STATE $c_0, \dots, c_n \gets 0$ \hfill \myComment{Counts for hash values}
\FOR{$j \in [D]$}
\STATE $v \gets \max\{\min\{h'_j, n\}, 0\}$ \hfill \myComment{Clip so that the value is between $0, n$}
\STATE $c_v \gets c_v + 1$ 
\ENDFOR
\STATE $c_{\geq 0}, \dots, c_{\geq n + 1} \gets 0$ \hfill \myComment{Counts for hash values above a certain threshold.}
\FOR{$r = n, \dots, 0$}
\STATE $c_{\geq r} \leftarrow c_{\geq r + 1} + c_r$
\ENDFOR
\FOR{$r \in [n]$}
    \STATE $\hphi_{\geq r} \gets c_{\geq r + 1} + \left(1 + \frac{p}{(1-p)^2}\right) c_r + \left(-\frac{p}{(1-p)^2}\right)c_{r - 1}$ 
\ENDFOR
\STATE Find $\bhn$ that minimizes $\|\bphi^{\bhn}_{\geq} - \hbphi_{\geq}\|_1$ using $\ell_1$-isotonic regression algorithm
\RETURN $\bhn$
\end{algorithmic}
\end{algorithm}

\subsection{Large Domain Case}\label{apx:hash-L1-loss-fast}

The large domain case is more complicated, as solving the optimization problem $\|\Gamma^B(\bhn) - \bphi_{\geq}^{\bhn^{\red}}\|_1$ does not seem to be as simple as isotonic regression. Nonetheless, we give a modified algorithm below that takes $\tO_\eps(D + n \log n)$ time and has a similar error bound (within a polylogarithmic factor). The idea is to use two noisy histograms in a similar manner as we did for $\ell_2^2$-error  (\Cref{alg:hash-L22-loss}): One noisy histogram (without hashing) is used to determine the high-value counts, whereas the other noisy histogram, with hashing, is used to determine the low-value counts.

\subsubsection{Computing $\Gamma^B$}

Recall from \Cref{obs:dependent-cut} that $\Gamma^B(\bn)_\ell$ does not depend on $\phi^{\bn}_{\geq r}$ for any $r > \ell$. Therefore, we may view $\Gamma^B(\bn)_\ell$ as a function of $\phi^{\bn}_{\geq 1}, \dots, \phi^{\bn}_{\geq \ell}$. We overload the notion and write $\Gamma^B(\phi^{\bn}_{\geq 1}, \dots, \phi^{\bn}_{\geq \ell})_\ell$ to denote $\Gamma^B(\bn)_\ell$ of a corresponding histogram $\bn$. We start by giving an algorithm for computing such a value. Recall that $\Gamma^B(\bn)_\ell$ is defined as the expected number of hash buckets whose total value is at least $\ell$. Let $\Xi_{\geq i}(\bn)$ denote the probability that a fixed hash bucket has total value at least $i$. By linearity of expectation, we have $\Gamma^B(\bn)_\ell = B \cdot \Xi_{\geq \ell}(\bn)$. Below, we keep updating $\Xi_{\geq i}$ as we include one more histogram value, using the fact that the hash function puts this histogram value in a random hash bucket.

\begin{algorithm}[h]
\caption{Computation of $\Gamma^B(\phi_{\geq 1}, \dots, \phi_{\geq \ell})_{\ell}$.} \label{alg:compute-gamma}
\begin{algorithmic}
\STATE {\bf Input: } $\phi_{\geq 1}, \dots, \phi_{\geq \ell}$, cumulative prevalence of a histogram  \\[2mm]
\STATE $\phi_{\geq \ell + 1} \gets 0$. \hfill \myComment{For notational convenience.}
\STATE $\Xi_{\geq 0} \gets 1$ and $\Xi_{\geq 1}, \dots, \Xi_{\geq \ell} \gets 0$ \hfill \myComment{Initialize $\Xi$ values}
\FOR{$r = 1, \dots, \ell$}
\FOR{$j=1, \dots, \phi_{\geq r} - \phi_{\geq r + 1}$}  
\STATE \vspace{-3.5mm}\hfill \myComment{Add a new histogram entry of value $r$}
\FOR{$k=\ell, \dots, 0$}
\STATE $\Xi_{\geq k} \gets \frac{B-1}{B} \cdot \Xi_{\geq k} + \frac{1}{B} \cdot \Xi_{\geq \max\{k - r, 0\}}$
\ENDFOR
\ENDFOR
\ENDFOR
\RETURN $B \cdot \Xi_{\geq \ell}$
\end{algorithmic}
\end{algorithm}

Since $\phi^{\bn}_{\geq 1} \leq n$, there are at most $n$ pairs $(r, j)$ considered by the algorithm. Therefore, the total runtime of \Cref{alg:compute-gamma} is $O(n \ell)$.

\subsubsection{Fast Post-processing Algorithm}

Our fast post-processing algorithm for large domains is stated below (\Cref{alg:hash-L1-loss-approx-fast-simplified}). For convenience, we use the convention that $\phi^{\bn}_{\geq 0} = n$.

\newcommand{\lb}{\mathsf{lb}}
\newcommand{\ub}{\mathsf{ub}}
\newcommand{\mi}{\mathsf{mid}}
\begin{algorithm}[h]
\caption{Efficient Anonymized Histogram Estimator w.r.t. $\ell_1$ loss, for large domains.}
\label{alg:hash-L1-loss-approx-fast-simplified}
\begin{algorithmic}
\STATE {\bf Input:} Two discrete Laplace-noised histograms $\tbh^{1}$ and $\tbh^{2, \red}$ given as
\begin{itemize}[nosep]
	\item $\tildeh^{1}_j \sim h_j + \DLap(p)$ (without any hashing), and
	\item $\tildeh^{2, \red}_j \sim h^{2, \red}_j + \DLap(p)$ where the random hash has $B$ buckets.
\end{itemize}
\STATE {\bf Parameter: } $m = \lceil 10 \log D / \log(1/p) \rceil$   \\[2mm]
\STATE Compute an estimate $\bhn^{1}$ of $\bn$ using \Cref{alg:L1-loss}
\STATE Compute an estimate $\bhn^{2, \red}$ of $\bn^{2, \red}$ using \Cref{alg:L1-loss}
\STATE $\hphi_{\geq 0} \gets n$
\FOR{$r \in [m]$}
\STATE $\lb = 0, \ub = \hphi_{\geq r - 1}$ \hfill \myComment{Binary search to find $\hphi_{\geq r}$}
\WHILE{$\ub > \lb + 1$}
\STATE $\mi \gets 
\lfloor (\ub + \lb) / 2 \rfloor$
\STATE $v \gets \Gamma^B(\hphi_{\geq 1}, \dots, \hphi_{\geq r - 1}, \mi)_{r}$ \hfill \myComment{See \Cref{alg:compute-gamma}}
\IF{$v \geq \phi^{\bhn^{2, \red}}_{\geq r}$}
\STATE $\ub \gets \mi$
\ELSE
\STATE $\lb \gets \mi$
\ENDIF
\ENDWHILE 
\STATE $v_{\ub} \leftarrow \Gamma^B(\hphi_{\geq 1}, \dots, \hphi_{\geq r - 1}, \ub)_{r}$
\STATE $v_{\lb} \leftarrow \Gamma^B(\hphi_{\geq 1}, \dots, \hphi_{\geq r - 1}, \lb)_{r}$
\IF{$|v_{\lb} - \phi^{\bhn^{2, \red}}_{\geq r}| \leq |v_{\ub} - \phi^{\bhn^{2, \red}}_{\geq r}|$}
\STATE $\hphi_{\geq r} \gets \lb$
\ELSE
\STATE $\hphi_{\geq r} \gets \ub$
\ENDIF
\ENDFOR
\STATE Find $\bhn$ that minimizes $\sum_{r=1}^m |\hphi_{\geq r} - \phi^{\bhn}_{\geq r}| + \sum_{r=m+1}^n |\phi^{\bhn^{1}}_{\geq r} - \phi^{\bhn}_{\geq r}|$ using $\ell_1$-isotonic regression algorithm
\RETURN $\bhn$
\end{algorithmic}
\end{algorithm}

\paragraph{Run Time Analysis.}
Recall that \Cref{alg:L1-loss} runs in $O(D + n \log n)$ time. In binary search, we call \Cref{alg:compute-gamma} for a total of at most $O(m \log n) = O(\log D \log n / \eps)$ times; since each call runs in $O(n r) = O(n \log D / \eps)$ time, the total running time of this step is $O(n \log n (\log D/\eps)^2)$. Finally, note that the last step is again an $\ell_1$-isotonic regression problem and therefore can be solved in $O(n \log n)$ time. Thus, in total the running time is $\tO_\eps(D + n)$ as desired. Note that we may assume without loss of generality that $D \leq n^{O(1)}$ as otherwise we may first hash to, e.g., $B' = O(n^2)$ buckets, which results in an increase in the expected $\ell_1$-error of at most $O(1)$.

\paragraph{Error Analysis.}
We now prove the error guarantee of the algorithm, which is stated precisely below.

\begin{theorem} \label{thm:l1-large-domain-fast}
Suppose that $B > 3nm$. Then, we have
\begin{align*}
\E[\|\bhn - \bn\|_1] \leq O(\sqrt{n^2 / B} \cdot \log n) + O(\sqrt{C_p B \log n}) + O(\sqrt{C_p n \log n}).
\end{align*}
\end{theorem}

Note that, since taking $p = e^{-\eps/4}$ makes the algorithm $\eps$-DP, we get the following:

\begin{corollary}
For all $\eps > 0$, \Cref{alg:hash-L1-loss-approx-fast-simplified} for $p = e^{-\eps/4}$ and $B = 10nm$ is $\eps$-DP, and achieves an expected $\ell_1$-error of $O(\sqrt{n} (\log (nD)) / \eps^{3})$.
\end{corollary}

The quantitative bound above matches \Cref{cor:hashing-err} to within a factor of $(\log n)^{1/4} / \eps^{0.5}$ (recall that we can assume without loss of generality that $D \le n^{O(1)}$).

To prove \Cref{thm:l1-large-domain-fast}, we will need the following auxiliary lemma, whose proof is given later in this section.

\begin{lemma} \label{lem:inverse-pointwise-apx}
Suppose that $B > 3nm$. Then,
\begin{align} \label{inverse-pointwise-apx-m}
\sum_{r=1}^m |\hphi_{\geq r} - \hphi^{\bn}_{\geq r}| \leq O\left(\|\Gamma^B(\bn) - \phi_{\geq}^{\bhn^{2, \red}}\|_1\right).
\end{align}
\end{lemma}

We are now ready to prove \Cref{thm:l1-large-domain-fast}.

\begin{proof}[Proof of \Cref{thm:l1-large-domain-fast}]
First, the expected $\ell_1$-error can be written as 
\begin{align*}
\E[\|\bhn - \bn\|_1] 
&= \E\left[\sum_{r=1}^n |\phi^{\bhn}_{\geq r} - \phi^{\bn}_{\geq r}|\right] 
\leq O\left(\E\left[\sum_{r=1}^m |\hphi_{\geq r} - \phi^{\bn}_{\geq r}| + \sum_{r=m+1}^n |\phi^{\bhn^{1}}_{\geq r} - \phi^{\bn}_{\geq r}|\right]\right),
\end{align*}
where the last step follows from the triangle inequality and our choice of $\bhn$.

We now plug in~\Cref{lem:inverse-pointwise-apx}, which yields:
\begin{align*}
\E[\|\bhn - \bn\|_1] &\leq O\left(\E\left[\|\Gamma^B(\bn) - \phi_{\geq}^{\bhn^{2, \red}}\|_1\right] + \E\left[\sum_{r=m+1}^n |\phi^{\bhn^{1}}_{\geq r} - \phi^{\bn}_{\geq r}|\right]\right) \\
&\overset{\text{\Cref{lem:hash-concen}, \Cref{thm:err-final}}}\leq O(\sqrt{n^2 / B} \cdot \log n + \sqrt{C_p B \log n}) + \E\left[\sum_{r=m+1}^n |\phi^{\bhn^{1}}_{\geq r} - \phi^{\bn}_{\geq r}|\right] %
\end{align*}
To bound the last term, we use \Cref{obs:cumulative-variance}, which gives:
\begin{align*}
&\E\left[\sum_{r=m+1}^n |\phi^{\bhn^{1}}_{\geq r} - \phi^{\bn}_{\geq r}|\right] 
\enspace~\leq~\enspace \sum_{r=m+1}^n \sqrt{\Var[\phi^{\bhn^{1}}_{\geq r}]} 
\enspace~\leq~\enspace \sum_{r=m+1}^n \sqrt{\kappa} \cdot \sqrt{\sum_{\ell=0}^n p^{|\ell - r|} \cdot \phi^{\bn}_{\ell}} \\
&\leq \sqrt{\kappa} \cdot \sqrt{\sum_{r=m+1}^n \frac{1}{r}} \cdot \sqrt{\sum_{r=m+1}^n r \cdot \left(\sum_{\ell=0}^n p^{|\ell - r|} \cdot \phi^{\bn}_{\ell}\right)} \quad \text{(Cauchy--Schwarz)}\\
&\leq \sqrt{\kappa} \cdot O(\sqrt{\log n}) \cdot \sqrt{\sum_{\ell=0}^n \phi^{\bn}_\ell \cdot \left(\sum_{r=m+1}^n r \cdot p^{|\ell - r|}\right)} \\
&\leq \sqrt{\kappa} \cdot O(\sqrt{\log n}) \cdot \sqrt{\left(\phi^{\bn}_0 \cdot \sum_{r=m+1}^n r \cdot p^{r}\right) + \left(\sum_{\ell=1}^n \phi^{\bn}_\ell \cdot \left(\sum_{r=1}^n r \cdot p^{|\ell - r|}\right)\right)}.
\end{align*}
The second term $\sum_{\ell=1}^n \phi^{\bn}_\ell \cdot \left(\sum_{r=1}^n r \cdot p^{|\ell - r|}\right)$ can be bounded in the exact same way as in \Cref{thm:err-final}, which gives an upper bound of $O(\sqrt{n/(1 - p)^2})$. The first term $\left(\phi^{\bn}_0 \cdot \sum_{r=m+1}^n r \cdot p^{r}\right)$ can be bounded as follows:
\begin{align*}
\phi^{\bn}_0 \cdot \sum_{r=m+1}^n r \cdot p^{r} &= D \left(p^{m + 1} \cdot \sum_{i=0}^n (i + m + 1)p^i\right) 
\enspace~\leq~\enspace D \left(\frac{p}{D^{10}} \cdot \left(\frac{m+1}{1 - p} + \frac{1}{(1 - p)^2}\right)\right) \\
&\leq O\inparen{\frac{1}{(1 - p)^2}}.
\end{align*}
Putting the four inequalities above together, we arrive at 
\begin{align*}
\E[\|\bhn - \bn\|_1] \leq O(\sqrt{n^2 / B} \cdot \log n + \sqrt{C_p B \log n}) + O(\sqrt{\kappa/(1 - p)^2}) \cdot O(\sqrt{n \log n}). &\qedhere
\end{align*}
\end{proof}

\begin{proof}[Proof of \Cref{lem:inverse-pointwise-apx}]
Let $\sigma := B/n$. We will prove by induction that 
\begin{align} \label{eq:inverse-pointwise-apx}
\sum_{r=1}^\ell |\hphi_{\geq r} - \hphi^{\bn}_{\geq r}| \leq 3\left(1 + \frac{3}{\sigma}\right)^\ell \left(\sum_{r=1}^\ell |\Gamma^B(\bn)_{r} - \phi_{\geq r}^{\bhn^{2, \red}}|\right),
\end{align} for all $\ell = 0, \dots, m$. The base case $\ell = 0$ is trivial. 

We will next prove the inductive step. Suppose that the bound \eqref{eq:inverse-pointwise-apx} holds for $\ell - 1$. We will now show that it also holds for $\ell$. Recall from \Cref{lem:lipschitz} that $\Gamma^B(\hphi_{\geq 1}, \dots, \hphi_{\geq \ell})_\ell$ is increasing in $\hphi_{\geq \ell}$ when $\hphi_{\geq 1}, \dots, \hphi_{\geq \ell - 1}$ are held fixed. This means that our binary search algorithm finds the optimum, which implies
\begin{align} \label{eq:binary-search-near-opt}
|\Gamma^B(\hphi_{\geq 1}, \dots, \hphi_{\geq \ell - 1}, \hphi_{\geq \ell})_{\ell} - \phi_{\geq \ell}^{\bhn^{2, \red}}| \leq |\Gamma^B(\hphi_{\geq 1}, \dots, \hphi_{\geq \ell - 1}, \phi^{\bn}_{\geq \ell})_{\ell} - \phi_{\geq \ell}^{\bhn^{2, \red}}|.
\end{align}
Applying the third inequality in \Cref{lem:lipschitz} repeatedly, we have
\begin{align}
&~|\Gamma^B(\phi^{\bn}_{\geq 1}, \dots, \phi^{\bn}_{\geq \ell - 1}, \hphi_{\geq \ell})_{\ell} - \phi_{\geq \ell}^{\bhn^{2, \red}}| \nonumber \\
(\text{\Cref{lem:lipschitz}}) \quad &~\leq~ |\Gamma^B(\hphi_{\geq 1}, \dots, \hphi_{\geq \ell - 1}, \hphi_{\geq \ell})_{\ell} - \phi_{\geq \ell}^{\bhn^{2, \red}}| + \frac{\sum_{r=1}^{\ell - 1} |\phi^{\bn}_{\geq r} - \hphi_{\geq r}|}{\sigma} \nonumber \\
(\text{From \eqref{eq:binary-search-near-opt}}) \quad &~\leq~ |\Gamma^B(\hphi_{\geq 1}, \dots, \hphi_{\geq \ell - 1}, \phi_{\geq \ell}^{\bn})_{\ell} - \phi_{\geq \ell}^{\bhn^{2, \red}}| + \frac{\sum_{r=1}^{\ell - 1} |\phi^{\bn}_{\geq r} - \hphi_{\geq r}|}{\sigma} \nonumber \\
(\text{\Cref{lem:lipschitz}}) \quad &~\leq~ |\Gamma^B(\phi^{\bn}_{\geq 1}, \dots, \phi^{\bn}_{\geq \ell - 1}, \phi^{\bn}_{\geq \ell})_{\ell} - \phi_{\geq \ell}^{\bhn^{2, \red}}| + \frac{2 \cdot \sum_{r=1}^{\ell - 1} |\phi^{\bn}_{\geq r} - \hphi_{\geq r}|}{\sigma} \nonumber \\
&~=~ |\Gamma^B(\bn)_{\ell} - \phi_{\geq \ell}^{\bhn^{2, \red}}| + \frac{2 \cdot \sum_{r=1}^{\ell - 1} |\phi^{\bn}_{\geq r} - \hphi_{\geq r}|}{\sigma}. \label{eq:diff-bound-1}
\end{align}
Again applying the second inequality in \Cref{lem:lipschitz}, we get
\begin{align*}
&|\hphi_{\geq \ell} - \phi^{\bn}_{\geq \ell}| \\
&\leq \frac{1}{1 - 1/\sigma} \cdot |\Gamma^B(\phi^{\bn}_{\geq 1}, \dots, \phi^{\bn}_{\geq \ell - 1}, \hphi_{\geq \ell})_{\ell} - \Gamma^B(\phi^{\bn}_{\geq 1}, \dots, \phi^{\bn}_{\geq \ell - 1}, \phi^{\bn}_{\geq \ell})_{\ell}| \\
&\leq \frac{1}{1 - 1/\sigma} \cdot\left(|\Gamma^B(\phi^{\bn}_{\geq 1}, \dots, \phi^{\bn}_{\geq \ell - 1}, \hphi_{\geq \ell})_{\ell} - \phi_{\geq \ell}^{\bhn^{2, \red}}| + |\Gamma^B(\phi^{\bn}_{\geq 1}, \dots, \phi^{\bn}_{\geq \ell - 1}, \phi^{\bn}_{\geq \ell})_{\ell} - \phi_{\geq \ell}^{\bhn^{2, \red}}| \right)\\
&\overset{\eqref{eq:diff-bound-1}}{\leq} \frac{1}{1 - 1/\sigma} \left(|\Gamma^B(\bn)_{\ell} - \phi_{\geq \ell}^{\bhn^{2, \red}}| + \frac{2 \cdot \sum_{r=1}^{\ell - 1} |\phi^{\bn}_{\geq r} - \hphi_{\geq r}|}{\sigma} + |\Gamma^B(\bn)_{\ell} - \phi_{\geq \ell}^{\bhn^{2, \red}}|\right) \\
&\leq 3|\Gamma^B(\bn)_{\ell} - \phi_{\geq \ell}^{\bhn^{2, \red}}| + \frac{3 \cdot \sum_{r=1}^{\ell - 1} |\phi^{\bn}_{\geq r} - \hphi_{\geq r}|}{\sigma} \qquad \inparen{\text{using }\ \frac{1}{1-1/\sigma} \le \frac{3}{2}}.
\end{align*}
Therefore, we have
\begin{align*}
\sum_{r=1}^{\ell} |\phi^{\bn}_{\geq r} - \hphi_{\geq r}| \leq 3|\Gamma^B(\bn)_{\ell} - \phi_{\geq \ell}^{\bhn^{2, \red}}| + \left(1 + \frac{3}{\sigma}\right) \cdot \left(\sum_{r=1}^{\ell - 1} |\phi^{\bn}_{\geq r} - \hphi_{\geq r}|\right).
\end{align*}
Plugging in the inductive hypothesis, we can conclude that \eqref{eq:binary-search-near-opt} also holds for $\ell$.

Finally, plugging in $\ell = m$ and using the fact that $\sigma \geq 3\ell$, we arrive at the claimed bound.
\end{proof}

\section{On Pan-Privacy}
\label{sec:pan-privacy-extension}

In the main body of our work, we only consider the notion of pan-privacy where, for every $t \in [n]$, the internal state of the algorithm after the $t$th step must be $\eps$-DP. This can be achieved for discrete Laplace-noised histogram as follows: start with $h_j$ drawn from $\DLap(p)$ for $p = e^{-\eps/2}$ for all $j \in [D]$. Then, at each step, increment the corresponding entry $h_j$.

While this algorithm suffices for our more relaxed notion, it does not satisfy the original notion of pan-privacy as defined in~\cite{DworkNPRY10}, which requires that, for every $t \in [n]$, both the internal state of the algorithm after the $t$th step \emph{and} the final output must be $\eps$-DP. A possible adaptation of the above algorithm to satisfy this notion of pan-privacy is to also add a noise drawn from $\DLap(p)$ to each entry of the histogram after the last element in the stream (before computing the final output). It is simple to see that this satisfies $\eps$-DP in the more restricted notion when we set $p = e^{-\eps/4}$.

Unfortunately, this adaptation does \emph{not} result in a discrete Laplace-noised histogram. Instead, the final histogram is noised by two i.i.d. discrete Laplace random variables (one from the initialization, and one from the final step). Due to this, we also have to adapt our estimation algorithm. Specifically, in \Cref{alg:L1-loss}, we replace $f$ by\footnote{Here $f * g$ denotes the convolution of $f$ and $g$, i.e., $(f * g)(j) = \sum_{i \in \Z} f(i) \cdot g(j - i)$ for all $j \in \Z$.} $f * g$ where $g$ is defined by
\begin{align*}
g(m) =
\begin{cases}
\frac{1+p^2}{(1 - p)^2} &\text{ if } m = 0, \\
-\frac{p}{(1 - p)^2} &\text{ if } m = -1 \text{ or } m = 1, \\
0 &\text{ if } m < -1 \text{ or } m > 1.
\end{cases}
\end{align*}
It is not hard to check that this results in an unbiased estimator, and an analogue of \Cref{thm:err-final} can be proved but with $C_p = O(1/(1 - p)^9)$. This in turn results in a worse error of $O(\sqrt{(n + D)\log n} / \eps^{4.5})$ instead of $O(\sqrt{(n + D)\log n} / \eps^{2.5})$ for the model considered in the main body. Other algorithms can be adapted similarly, again with worse dependency of $\eps$ in the error bounds.
\end{document}